\documentclass[11pt,a4paper]{article}
\pdfoutput=1
\usepackage{jinstpub}

\title{Multi-Grid Detector for Neutron Spectroscopy: Results Obtained on Time-of-Flight Spectrometer CNCS}

\author[a]{M.~Anastasopoulos,}
\author[a]{R.~Bebb,}
\author[b]{K.~Berry,}
\author[c]{J.~Birch,}
\author[a]{T.~Bry\'s,}
\author[d]{J.-C.~Buffet,}
\author[d]{J.-F.~Clergeau,}
\author[a]{P.~P.~Deen,}
\author[e]{G.~Ehlers,}
\author[d]{P.~van Esch,}
\author[b]{S.~M.~Everett,}
\author[d]{B.~Guerard,}
\author[a,f]{R.~Hall-Wilton,}
\author[g]{K.~Herwig,}
\author[c]{L.~Hultman,}
\author[a,c]{C.~H\"oglund,}
\author[a]{I.~Iruretagoiena,}
\author[a]{F.~Issa,}
\author[c]{J.~Jensen,}
\author[a,1]{A. Khaplanov, \note{Corresponding author.}}
\author[a,h]{O.~Kirstein,}
\author[a]{I.~Lopez~Higuera,}
\author[a]{ F.~Piscitelli,}
\author[a]{ L.~Robinson,}
\author[a,c]{S.~Schmidt,}
\author[a]{I.~Stefanescu,}

\affiliation[a]{European Spallation Source, P.O Box 176, SE-22100 Lund, Sweden}
\affiliation[b]{Instrument and Source Division, Spallation Neutron Source, 1 Bethel Valley Road, Oak Ridge, TN 37831-6476, USA}
\affiliation[c]{Link\"{o}ping University, Thin Film Physics division, IFM, SE-581 83 Link\"{o}ping, Sweden}
\affiliation[d]{Institute Laue Langevin, 71 avenue des Martyrs, FR-38042 Grenoble, France}
\affiliation[e]{Quantum Condensed Matter Division, Spallation Neutron Source, 1 Bethel Valley Road, Oak Ridge, TN 37831-6475, USA}
\affiliation[f]{Mid-Sweden University, SE-85170 Sundsvall, Sweden}
\affiliation[g]{Instrument and Source Division, Spallation Neutron Source, 1 Bethel Valley Road, Oak Ridge, TN 37831-6466, USA}
\affiliation[h]{School of Mechanical Engineering, University of Newcastle, Callaghan, Australia}

\emailAdd{Anton.Khaplanov@esss.se}

\abstract{The Multi-Grid detector technology has evolved from the proof-of-principle and characterisation stages. Here we report on the performance of the Multi-Grid detector, the MG.CNCS prototype, which has been installed and tested at the Cold Neutron Chopper Spectrometer, CNCS at SNS. This has allowed a side-by-side comparison to the performance of $^3$He detectors on an operational instrument. The demonstrator has an active area of 0.2 m$^2$. It is specifically tailored to the specifications of CNCS. The detector was installed in June 2016 and has operated since then, collecting neutron scattering data in parallel to the He-3 detectors of CNCS. 
In this paper, we present a comprehensive analysis of this data, in particular on instrument energy resolution, rate capability, background and relative efficiency. Stability, gamma-ray and fast neutron sensitivity have also been investigated. The effect of scattering in the detector components has been measured and provides input to comparison for Monte Carlo simulations. All data is presented in comparison to that measured by the $^3$He detectors simultaneously, showing that all features recorded by one detector are also recorded by the other. The energy resolution matches closely. We find that the Multi-Grid is able to match the data collected by $^3$He, and see an indication of a considerable advantage in the count rate capability. Based on these results, we are confident that the Multi-Grid detector will be capable of producing high quality scientific data on chopper spectrometers utilising the unprecedented neutron flux of the ESS.}


\begin{document}
\maketitle

\section{Introduction}

The Multi-Grid detector~\cite{cite:khaplanov1, cite:correathes} for thermal and cold neutrons has been introduced at ILL and developed by a collaboration between ILL, ESS and Link\"{o}ping University in order to provide an alternative to He-3 detectors for applications where the cost and availability of this rare isotope would be prohibitive~\cite{cite:he31, cite:he32, cite:he33}. This is in particular true for large-area detectors needed at time-of-flight (ToF) neutron spectrometers, such as CSPEC, T-REX and VOR that will be built for the ESS~\cite{cite:cdr, cite:tdr, cite:vor, cite:kirstein}. The detector exploits the neutron conversion reaction in $^{10}$B in order to detect neutrons. It combines the wire proportional chamber readout with thin films of enriched boron carbide in order to detect thermal neutrons. The geometry and the readout of the detector are designed in such a way that the interaction position of each neutron can be localised to a voxel of typically (20$\times$20$\times$10)~mm$^3$. Combined with the timestamp assigned to each neutron, the detector is able to record time-dependent maps of neutrons scattered by a sample. In this way, wide ranges of momentum and energy transfer can be studied in a single measurement~\cite{cite:cncs, cite:in5, cite:let, cite:4seasons}. 

In this paper, we look in depth at the performance of the Multi-Grid detector at the ToF spectrometer CNCS. Several sets of measurements with vanadium and crystalline samples have been performed. These allow us to benchmark the performance of the detector in ToF operation, determine the achievable energy resolution, as well as get a handle on the efficiency of the detector. Detailed investigations of the shape of the measured energy spectrum have allowed us to investigate the impact of neutron scattering events within the detector. A measurement with a single crystal was used to test and compare the Multi-Grid and $^3$He detector performance at a high local instantaneous rate. We have also identified fast neutron detections and were able to perform tentative comparisons of sensitivities of the two technologies to fast neutrons. Backgrounds and long-term stability could also be measured. Several of these measurements are only possible on a ToF instrument and are not accessible in lab source or continuous beam tests that have been done previously. This experiment follows an earlier test on the IN6 spectrometer at the ILL, where a Multi-Grid prototype was operated in the instrument for 2 weeks\cite{cite:in6test}. The current study is the first where an Multi-Grid detector has been operated continuously for over half a year in constant conditions and no access to the detector, other than via the remote retrieval of data. 

\section{The Test of the Multi-Grid Detector}

The Multi-Grid technology has previously been extensively characterised in test setups~\cite{cite:khaplanov1, cite:correathes, cite:in6test, cite:correa, cite:guerard, cite:hoglund}, however, a test at an operational instrument was much preferred in order to fully compare the new technology to the more familiar $^3$He. A test was agreed to be performed at CNCS at SNS in Oak Ridge National Laboratory, for which a new Multi-Grid detector was designed and installed on the existing instrument directly adjacent to the current $^3$He detectors. This allowed measurements in parallel with the existing $^3$He detectors over an extended period of time -- primarily during user experiments, but also to perform dedicated tests. 

\subsection{CNCS}

\begin{figure}[tbp] 
\centering
\includegraphics[width=0.99\textwidth]{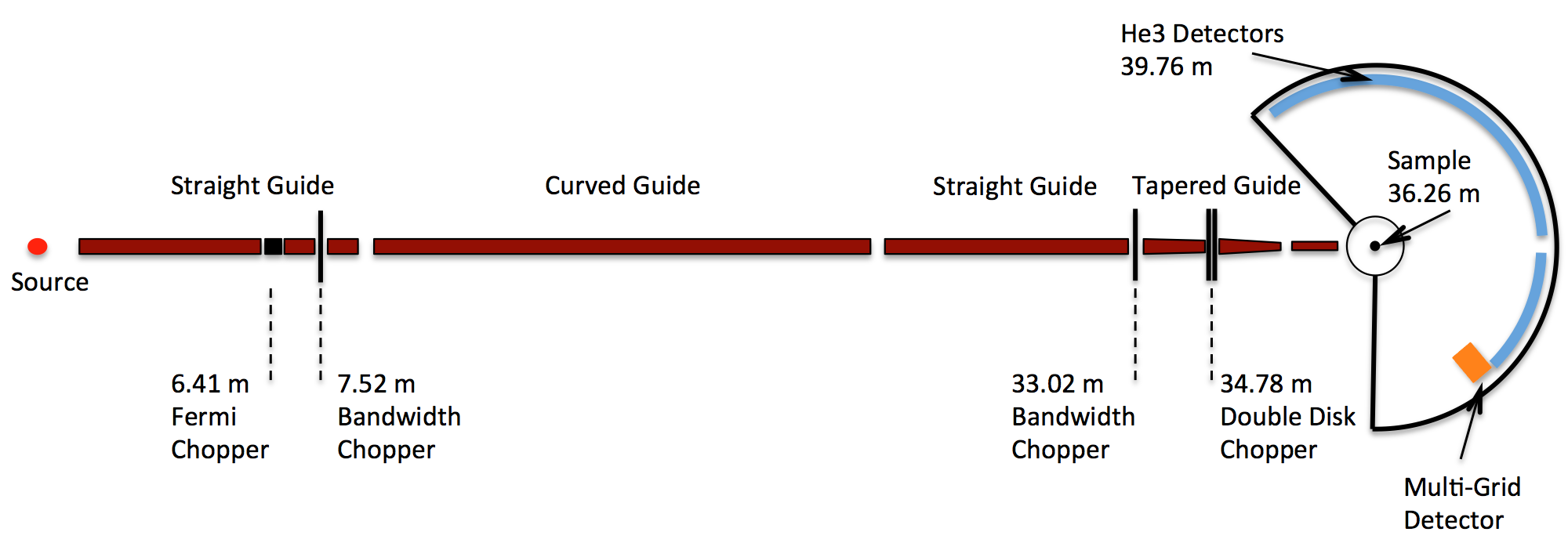}
\caption{Schematic layout of CNCS at SNS~\cite{cite:cncs, cite:cncs2}, indicating distances from the source to the main components and the location of the Multi-Grid detector.}
\label{fig:cncs}
\end{figure}

The Cold Neutron Chopper Spectrometer, CNCS, is the instrument at the SNS designed for inelastic studies at low energies up to typically 20~meV, with energy resolution down to $dE/E=1\%$. The instrument specialises in magnetism, structural excitations and dynamics in confined geometries~\cite{cite:cncs, cite:cncs2}. The layout of the instrument is shown in figure~\ref{fig:cncs}. Neutrons are extracted from a cold coupled moderator and are delivered to the sample position 36.25~m away via a neutron guide and a series of choppers. A detector array composed of 400 one-inch 2-m long $^3$He tubes with a total area of 20~m$^2$ is placed in a cylindrical geometry at 3.5~m from the sample. This makes it possible to detect scattered neutrons over a large range of scattering angles, allowing a large range of momentum transfers, $Q$, to be investigated alongside with the measurement of the final energy of the scattered neutrons, thereby gaining an understanding of the energy transfer from the neutron to the sample and vice versa. The chamber that encloses the sample position and the flight path of the scattered neutrons to the detectors is filled with argon at atmospheric pressure. This minimises the probability of neutrons scattering a second time before reaching the detector, while avoiding the complexity of a vacuum tank. 

CNCS employs $^3$He tubes to detect the scattered neutrons. The 2~m long tubes, filled with $^3$He at 6~bar, are position-sensitive. They are equipped with a resistive anode wire which is readout at both ends, allowing the reconstruction of the interaction position. The tubes are virtually divided into 128 pixels to which every neutron is assigned after processing the signals. Additionally, each detected neutron is time-stamped and stored by the data acquisition system. 

The scattering chamber of CNCS has a large amount of vacant space, making a future upgrade to twice the detector coverage possible. Presently, we take advantage of some of this space for the test of the Multi-Grid detector. The position where it was installed is indicated in figure~\ref{fig:cncs}.

\subsection{Multi-Grid Detector for the CNCS Test, MG.CNCS}

\begin{figure}[tbp] 
\centering
\includegraphics[width=0.45\textwidth]{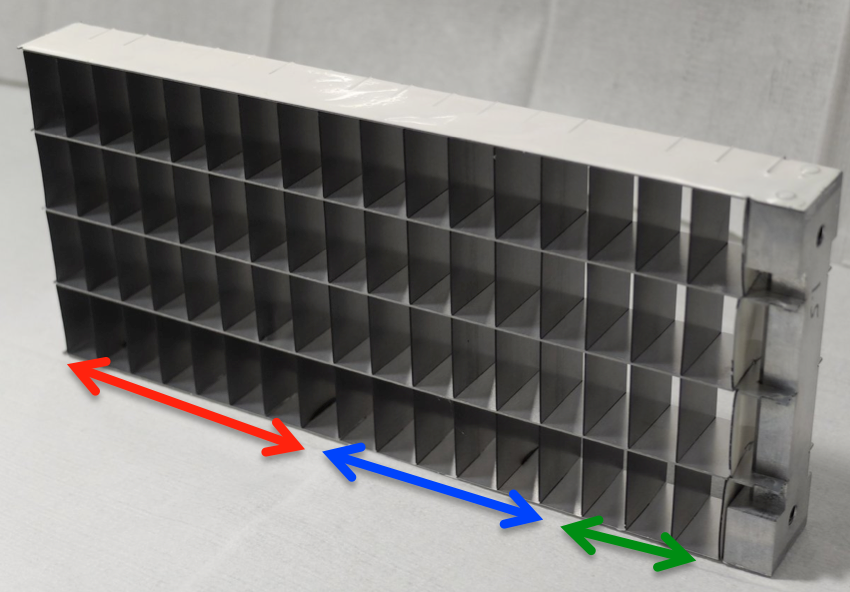}
\caption{A grid. The three regions of different thicknesses of boron layers are shown by the arrows. The coated blades are the vertical components in the image. The length of the grid is approximately 20~cm.}
\label{fig:grid}
\end{figure}

\begin{figure}[tbp] 
\centering
\includegraphics[width=0.5\textwidth]{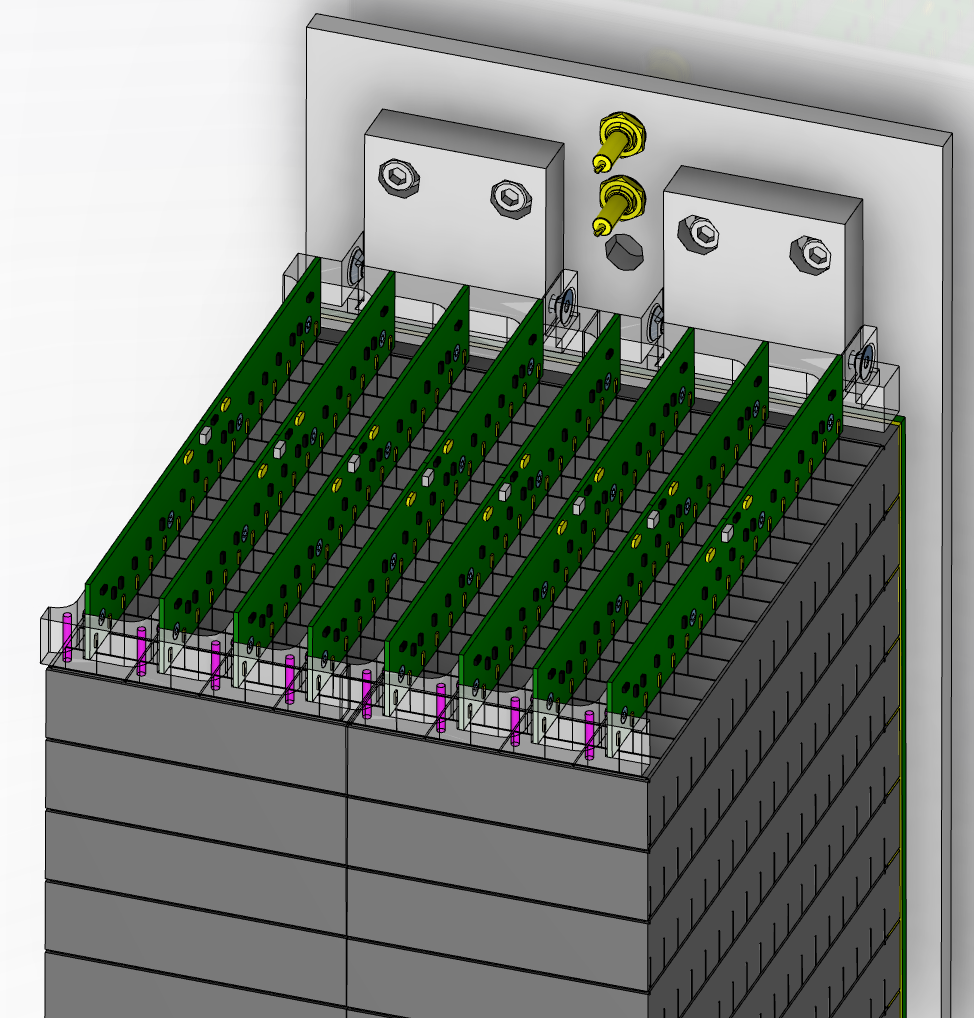}
\includegraphics[width=0.181\textwidth]{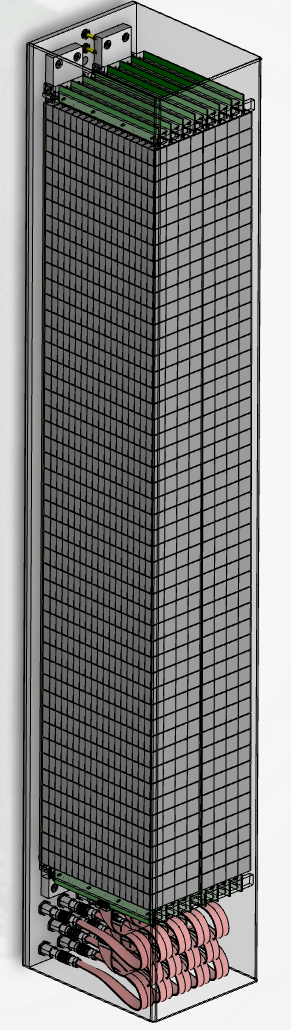}
\includegraphics[width=0.25\textwidth]{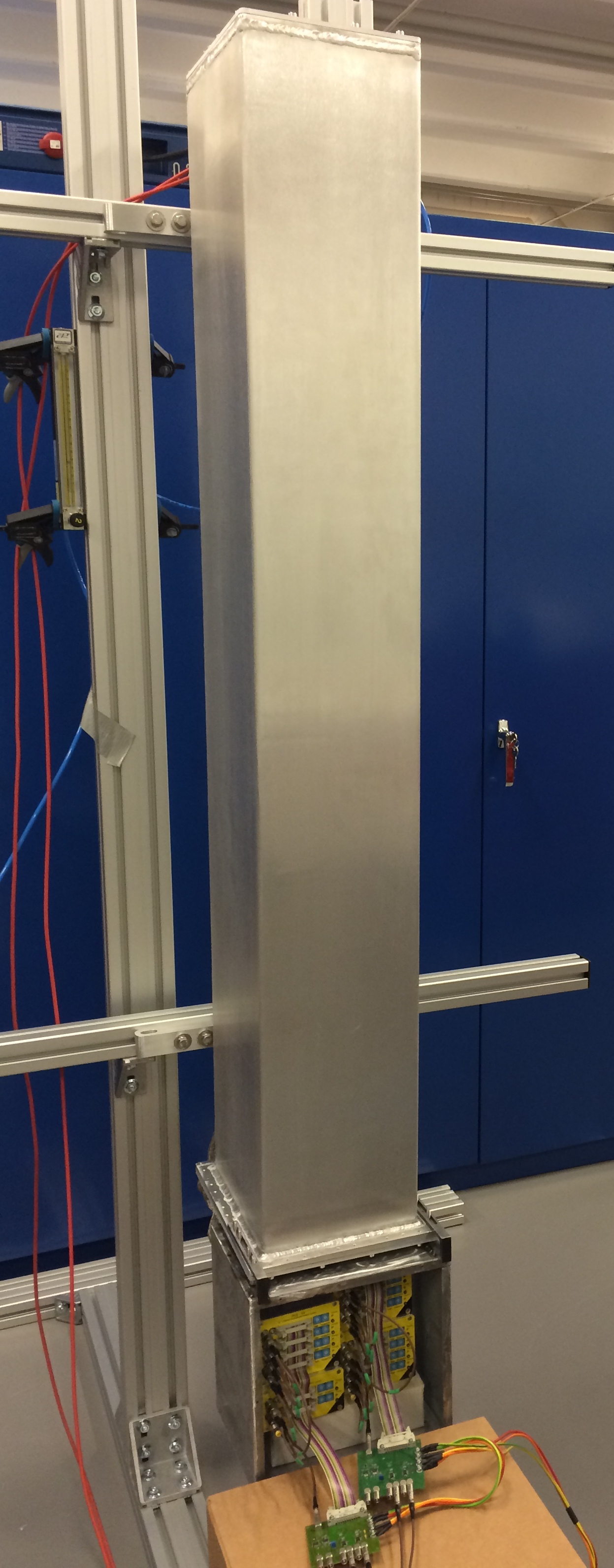}
\caption{\textbf{Left and middle:} CATIA6 design of MG.CNCS. \textbf{Right:} completed detector on a test stand. The height of MG.CNCS is approximately 1.4~m.}
\label{fig:mgcncs}
\end{figure}

The detector is constructed from grids, such as one shown in figure~\ref{fig:grid}. The elements of the grids crossed orthogonally by incoming neutrons are called \emph{blades}. These are coated with $^{10}$B$_4$C, the neutron-sensitive medium of the detector. The other elements of the grids are not coated. Grids are stacked with no electrical contact from grid-to-grid, forming modules, where the number of grids is chosen in order to cover the necessary height of the detector. The width of the detector is tiled by placing multiple modules side-by-side. The MG.CNCS detector is shown in figure~\ref{fig:mgcncs}. The 2 modules are placed alongside each other, each 48 grids high, resulting in a sensitive area of about 0.2~m$^2$. 

The grids are insulated from each other and function as cathodes of the detector. The anodes are wires that are passed through all of the grids in each stack. The cells of the grids thus form rectangular proportional gas chambers, whose cathodes are segmented in length. By detecting signals on both grids and wires and finding coincidences, the position of detection of a neutron can be localised to within a voxel equal to the size of the grid cell, in the present design, ($22 \times 22 \times 11$)~mm$^3$ (W$\times$H$\times$D). The detector is filled with a mixture of Ar (80\%) and CO$_2$ (20\%) at atmospheric pressure. 

\begin{figure}[tbp] 
\centering
\includegraphics[width=0.49\textwidth]{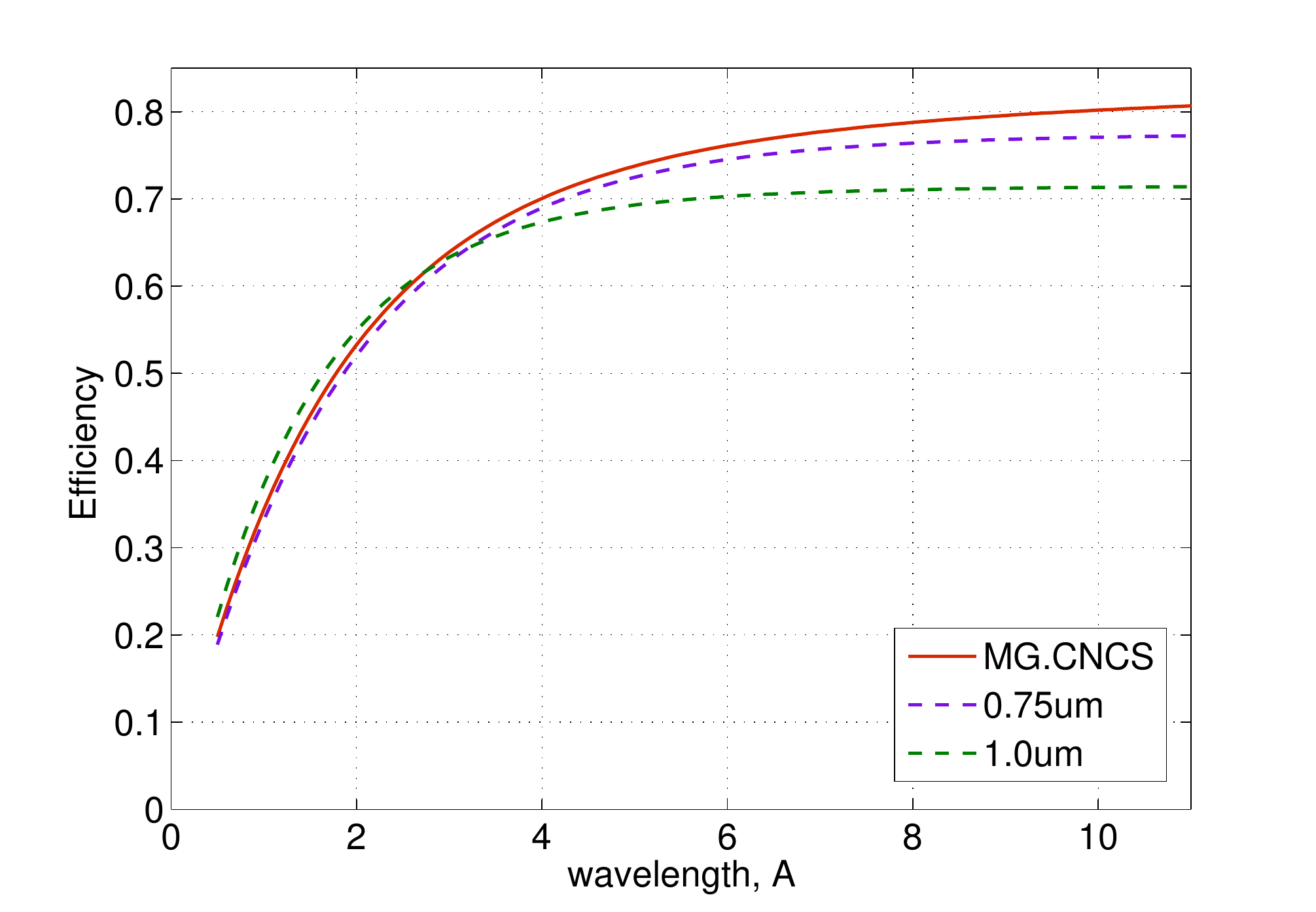}
\includegraphics[width=0.49\textwidth]{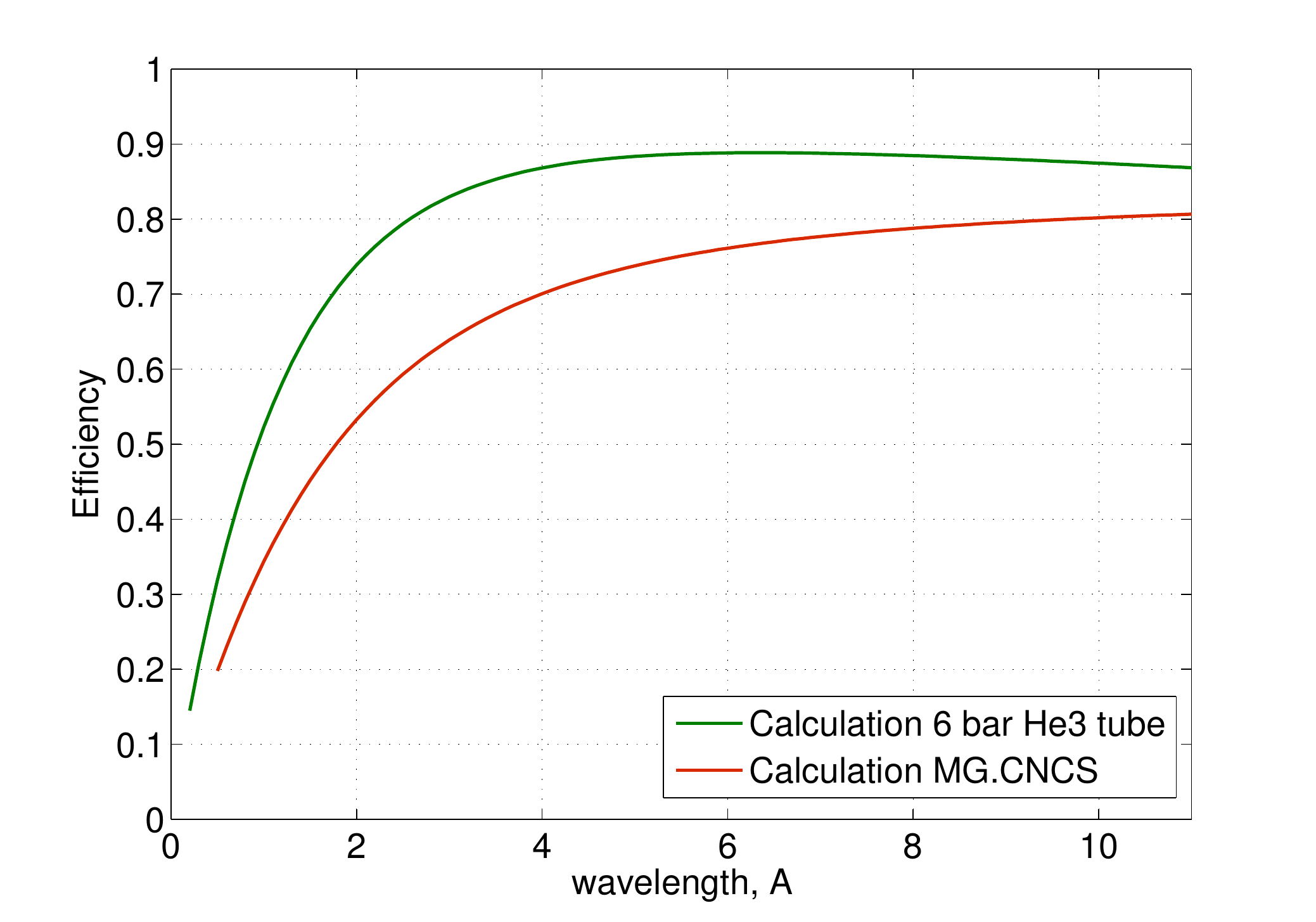}
\caption{\textbf{Left:} Efficiency calculated for three choices of boron layer thicknesses. The MG.CNCS configuration with 0.5, 1.0 and 1.5~$\mu$m layers is compared to constant thicknesses of 0.75 and 1.0~$\mu$m. \textbf{Right:} The 3-thickness configuration of MG.CNCS compared to calculated efficiency for 1-inch tubes filled with 6~bar of $^3$He.}
\label{fig:effopt}
\end{figure}

The wavelength-dependent efficiency of a multi-layer $^{10}$B detector depends on the number and the thicknesses of the layers used. The highest efficiency is reached when the thickness of the layers increases from the front to the back of the grid. 16 double layers were chosen and the layer thicknesses were optimised~\cite{cite:piscitelli} so that the efficiency is maximised for a spectrum centred on 4~\AA, with the constraint that no more than 3 different layer thicknesses are used. The resulting configuration is 7 blades with $0.5\mu m$ coating at the front of each grid, 7 blades with $1\mu m$ in the middle, and 3 blades with $1.5\mu m$ at the back. Note that a 1-side coated blade is used front- and rear-most in each grid. Figure~\ref{fig:effopt} shows the efficiency for the chosen configuration compared to constant layer thicknesses of 0.75 and 1.0~$\mu$m and to the calculated efficiency of the 6-bar $^3$He tubes of the CNCS detector array. In these plots, efficiency is shown taking into account the sensitive area of the detecting elements, \emph{i.e.} grids of the Multi-Grid and tubes of the $^3$He. The gaps between grids and tubes are excluded.

A very low background is one of the main requirements of a neutron spectrometer, since signals are 3 to 4 orders of magnitude below the intensity of the peak flux. In order to reduce the internal background of the demonstrator, high-purity aluminium was used for the blades coated with $^{10}$B$_4$C. It has been shown previously, that this grade of aluminium has about 2 orders of magnitude lower concentration of $\alpha$-emitting impurities (mainly U and Th)~\cite{cite:alpha}. Alpha-particle emissions from the grid materials contribute directly to background in the same energy range as neutron conversions. Therefore we expect 2 orders of magnitude better background than in case of standard Al, as for example in our test at the IN6 spectrometer at the ILL~\cite{cite:in6test}. The non-coated elements of the grids were manufactured from a standard Al (5000-series) and plated with 25~$\mu$m of Ni. This layer is sufficient to stop all $\alpha$-particles emitted from Al underneath~\cite{cite:alpha}.

\begin{figure}[tbp] 
\centering
\includegraphics[width=0.4\textwidth]{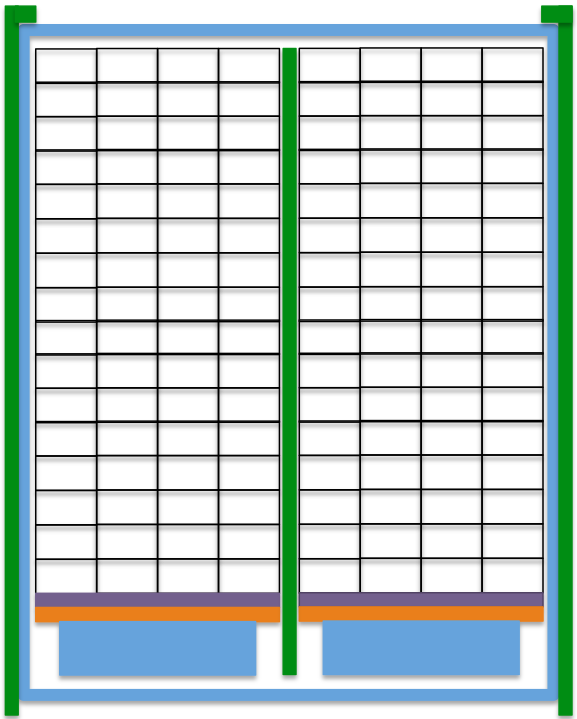}
\caption{Sketch of MG.CNCS seen from the top, showing the location of the Gd$_2$O$_3$ (purple) and B$_4$C (green) shielding. Al parts are shown in blue and PCBs in orange. Neutrons arrive from the top in this figure.}
\label{fig:shielding}
\end{figure}

In order to minimise the effect of scattering within the detector several highly absorbing materials were used. Neutron shielding based on Gd$_2$O$_3$ was used immediately after the last boron-coated blade. B$_4$C flexible shielding was used between the two grid modules, as well as on all outer surfaces of the detector vessel, except the neutron window. Figure~\ref{fig:shielding} shows a sketch of the location of the shielding materials. The Gd$_2$O$_3$ layers are 1~mm thick and B$_4$C shielding is 2~mm thick.

\subsection{Readout}

The electronics used to readout the MG.CNCS are divided into two parts -- Front-End Electronics (FEE) and Data Acquisition (DAQ). The FEE is located immediately adjacent to the detector in order to minimise the length of the cables carrying non-amplified signals. MUX modules, commercially available from Mesytec GmbH~\cite{cite:mesytec}, were used as the FEE. These modules are comprised of 16-channel boards including pre amplifiers, shapers, discriminators and a multiplexing stage. Several modules can be connected in a chain. Two such chains were used -- one for all the grids (96 channels) and one for all wires (128 channels). Each chain can trigger on any single channel and will accept one more channel triggered simultaneously (\emph{i.e.} multiplicity 1 or 2). The output signals are then two amplitudes and two signals that encode the numbers of the two channels that triggered during that event. With two sets of such units, the entire detector can be readout via 8 channels (2 wire amplitudes, 2 grid amplitudes, 2 wire positions and 2 grid positions). 

The possibility to record multiplicity 2 in each events for both wires and grids can be used in two ways. For grids, many events trigger 2 grids, since a charge generated in any cell can be sensed in the neighbouring cell of an adjacent grid. As the amplitudes of signals from both grids are recorded, it is possible to enhance the position reconstruction to that better than the height of one grid. While this improvement in resolution has not been necessary in the current detector (the height of a grid matches the required resolution), this gives us data for future investigation of possible gain in resolution. Multiplicity 2 in wires does not generally occur in the same event, since wires run through long tubes and are not influenced by electric fields in neighbouring tubes. Therefore, in most events, wire 2 signal is not active. When it is, it indicates the presence of random coincidences, and gives us information on their frequency. 

\begin{figure}[tbp] 
\centering
\includegraphics[width=0.99\textwidth]{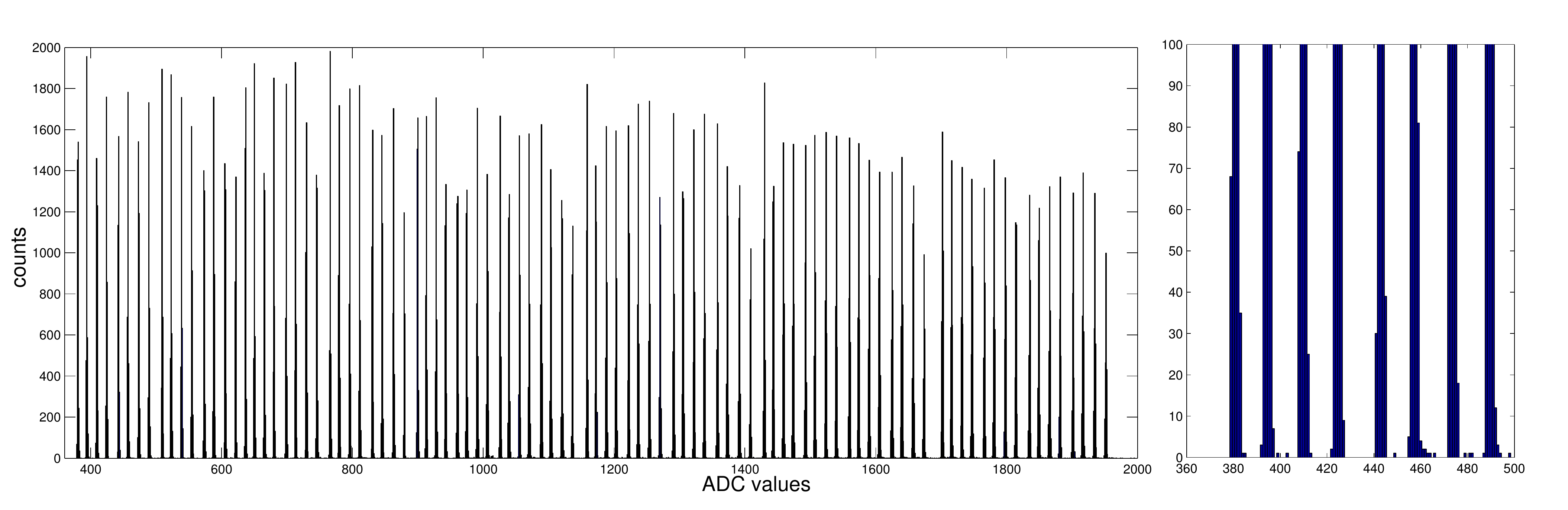}
\caption{\textbf{Left:} Spectrum of the \emph{grid position} signal. The spectrum is an analogue channel encoding generated by the MUX front-end units; here, the signal has been digitised by MADC-32 module. \textbf{Right:} a magnification of the bases of the first 8 peaks, indicating that channels are well separated.}
\label{fig:gridpos}
\end{figure}

The 8 signals are led to the DAQ and digitised. Two DAQ systems are being used. One is a Mesytec MADC-32 module, a 32-channel peak sensing ADC. The other is FastComTec MCA4~\cite{cite:fastcomtec}. The MADC digitises the 8 detector signals synchronously, using the trigger provided by the FEE. For each event, the 4 amplitudes and 4 positions are stored along with a time stamp. In the case of the MCA4 DAQ, only 4 digitisation channels are available. Out of the 8 signals, the 4 chosen are (1) wire 1 amplitude, (2) wire 1 position, (3) grid 1 position, (4) grid 2 position. With this selection, the energy of the event is available through wire 1 amplitude; the voxel can be reconstructed. Figure~\ref{fig:gridpos} shows an example of a position spectrum for grids. The counts between the peaks drop off to zero, providing a good separation of individual channels.

\subsection{Installation at CNCS}

The detector was transported to the SNS and installed during June 2016. The detector was mounted onto the existing detector support structure of the CNCS chamber. The position was then adjusted so that the detector was vertical and the entrance window was facing the sample direction. Laser-based surveying equipment was used to make fine adjustments to the position and the orientations, as well as to determine the position of the detector with respect to the sample position. We defined the sample centre position as origin, measured coordinates of a reference position on the detector and calculated the positions of each detector cell based on the CAD model in CATIA6. In this way we obtained a complete information about the position and distance to the sample for each voxel.

\begin{figure}[tbp] 
\centering
\includegraphics[width=0.49\textwidth]{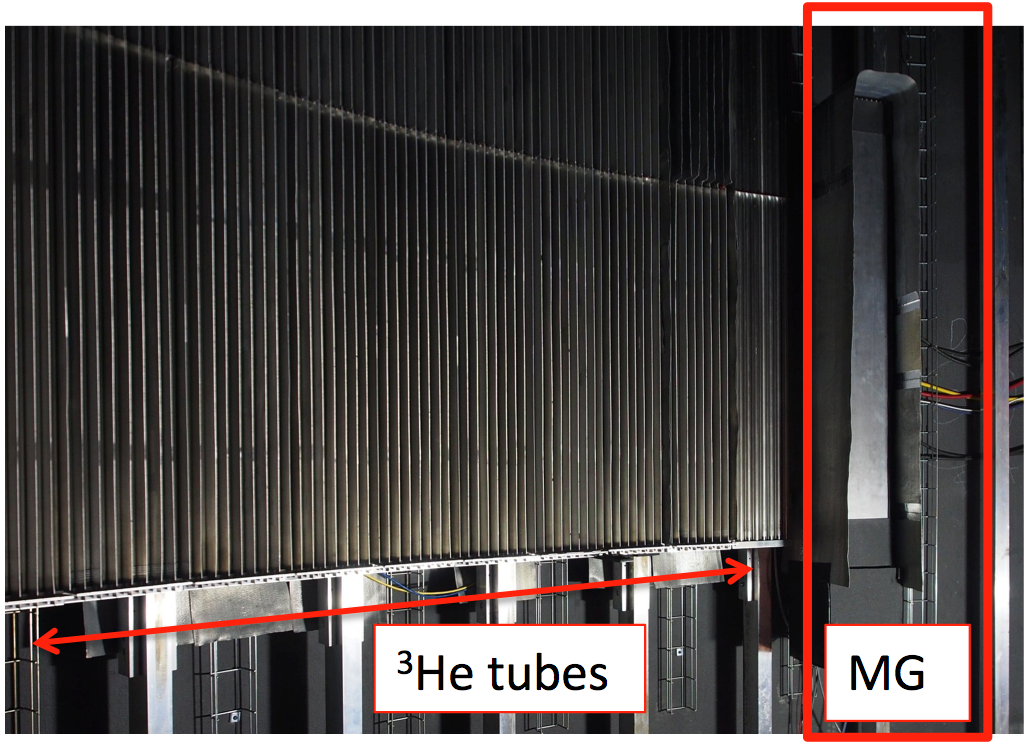}
\includegraphics[width=0.49\textwidth]{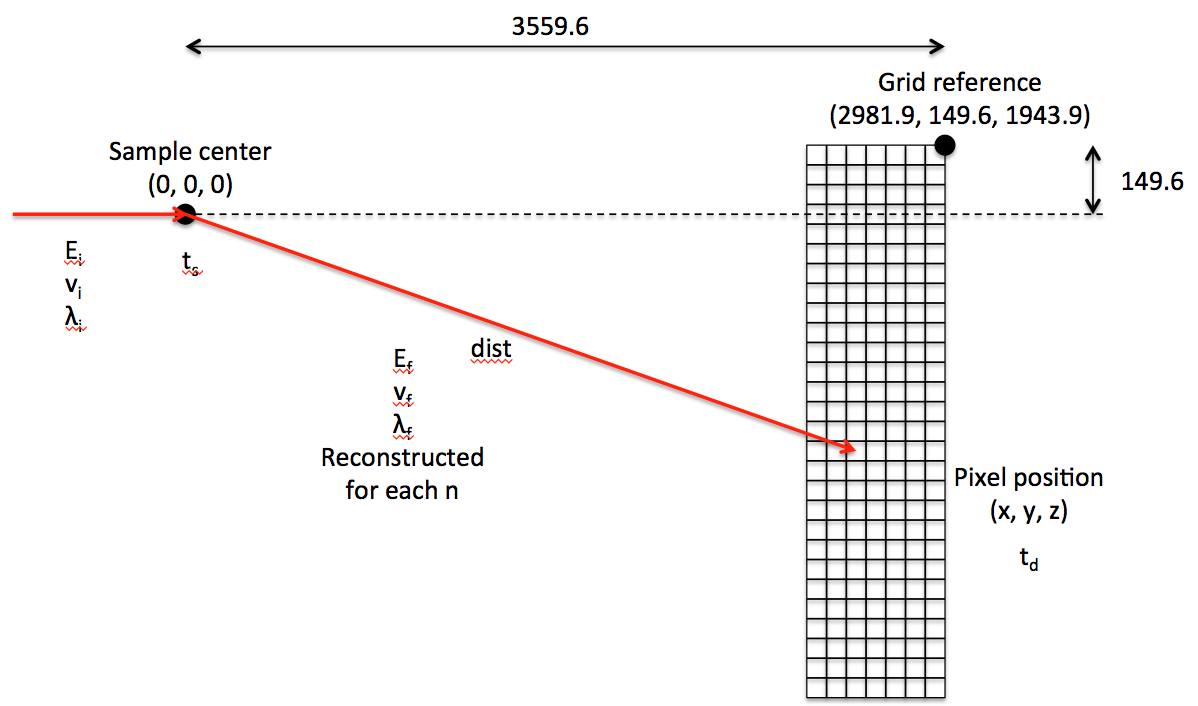}
\caption{\textbf{Left:} photo of the Multi-Grid installed at the right-most edge of the $^3$He detector array. \textbf{Right:} schematic showing the position of the detector with respect to the sample. Distances are shown in mm.}
\label{fig:installed}
\end{figure}

The final position of the detector was at 56.9$^\circ$ scattering angle in the horizontal plane. The top of the detector is 15 cm above the scattering plane and the bottom is 95 cm below the scattering plane, approximately at the same level as the lower ends of the $^3$He tubes. In this position, the detector sees a range of incoming neutron angles similar to that of the $^3$He tubes. The detector in its final position is shown in figure~\ref{fig:installed}.

The FEE enclosure is attached to the bottom of the detector vessel, while the data acquisition is setup outside of the chamber. Cabling of 12 meters length was used to connect the FEE to the data acquisition setup. The gas supply and exhaust were connected via a flexible gas lines to the detector. The gas flow regulator is situated outside the chamber and supplies gas at a rate such that the volume of the detector is refreshed twice per day. All necessary cables and the gas line were fed through a 2-inch wall penetration, that was subsequently sealed to maintain the Ar atmosphere inside the chamber. 

Additionally, a 3$\times$3-inch NaI scintillation detector was installed close to the Multi-Grid detector. This detector was enveloped in borated shielding to minimise the impact of neutrons, and is used to measure the $\gamma$ flux. Similarly to the Multi-Grid detector, its data can be time-stamped. This allows to measure ToF spectra of $\gamma$ rays.

\subsection{Commissioning of the detector}

A number of measurements were done at the end of the installation in order to verify the operation of the detector and tune the FEE and DAQ. These were performed using a moderated neutron source ($^{252}$Cf), and $\gamma$ sources ($^{137}$Cs and $^{241}$Am). The neutron source was used to verify that all channels were working and the connections to the DAQ were reliable. The combinations of neutron and $\gamma$ source measurements allowed to set up the bias voltage and discriminator thresholds, using the methods described in~\cite{cite:khaplanov2}. The thresholds were set so that the $\gamma$-ray signal is just visible in the raw pulse hight spectra. This allowed to set the final threshold in the software and allows to fine-tune it at any time during the measurements. In a typical measurement, on the order of 10\% of events are rejected by the software threshold. In addition to its flexibility, this approach also allows us to test whether or not any features in the measured spectra may be due to gamma rays. The source tests were completed before the scattering chamber was sealed and filled with Ar. Once the neutron production restarted at the SNS, a series of beam tests was carried out. These were a set of vanadium sample measurements at 4 incoming energies, allowing final adjustments of the DAQ. Following these, the detector was no longer accessed other than for retrieval of data from the DAQ system, until the time of the writing, approximately 8 months later.

\section{Measurements}

\begin{figure}[tbp] 
\centering
\includegraphics[width=0.99\textwidth]{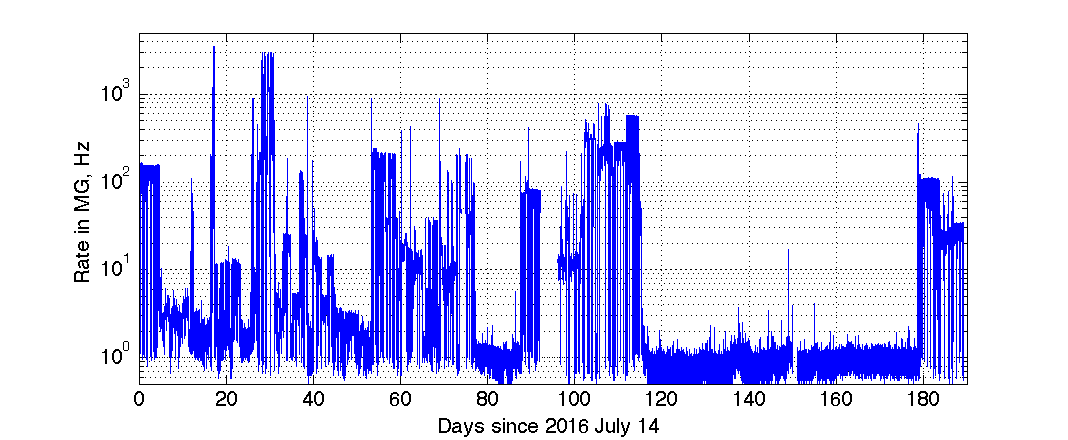}
\caption{Time average count rate measured in MG.CNCS as a function of time. Data are shown for the first half year of operation.}
\label{fig:statistics}
\end{figure}

The Multi-Grid detector has been operating continuously since the commissioning in July 2016. The bulk of the measurement time is taken by user experiments performed at CNCS. Additionally, it has been possible to perform several dedicated measurements. These latter provide the majority of the characterisation results presented in this section. For dedicated experiments, data from the full $^3$He array and sample conditions are available for comparison and analysis. Figure~\ref{fig:statistics} shows the overall rate recorded by the MADC digitiser since the beginning of the experiment. The varying rate indicates changes of the settings of the beam and sample environment for the needs of the experiments being performed. The low rates during the first 60 days correspond to a sample environment that blocks the view of the side of the detector array where the Multi-Grid is positioned. At days 80-90, there was a scheduled maintenance shutdown, while from day 120 the beam was down for the Christmas shutdown. These periods contribute to our background measurement. There are several interruptions in the plot -- twice for about a day (at about day 75 and 150) and once for 3 days (at about day 95) -- when the data acquisition was down. Data for these periods is available from the MCA4 system. The detector has been powered continuously. 

The $^3$He detector array is composed of 400 tubes which are grouped into banks of 8, so called \emph{8-packs}. In the following, the majority of the data is shown alongside the data from the nearest bank, \#50. In some cases, the comparison includes only the bottom half of bank \#50, in order to select the area as close as possible to the Multi-Grid. The data from user experiments is also compared to the data from the closest 8-pack, however, the detailed information on the sample properties and sample environment conditions is not available, as this data belongs to the principal investigators of the experiments. Nevertheless, it is useful to analyse the response of the detector to the widely varying neutron rates, as well as to monitor the stability of the detector over time. Furthermore, the downtimes of the accelerator provide valuable opportunities to measure background, or \emph{dark counts}. 

For all figures where the Multi-Grid detector spectra are compared to $^3$He detectors, the spectra are shown as measured. The acquisition times are equal. The counts have been normalised to the solid angle taken by the Multi-Grid relative to that of $^3$He bank \#~50 (this ratio is 0.508).

\subsection{Vanadium Measurements}

The workhorse neutron scattering sample for characterisation and calibration measurements is vanadium. Natural vanadium is composed to 99.75\% of the $^{51}$V isotope, whose scattering cross-section is nearly purely incoherent. This means that a vanadium sample (a small rod of metallic vanadium placed in a standard vanadium sample container) exposed to the neutron beam results in a uniform scattering pattern in 4$\pi$. The deviation from uniformity in the intensity of the detected neutrons as a function of scattering angle is then a function of the geometry of the instrument setup. This includes several angle-dependent effects, such as absorption and double scattering in the sample, sample holder, sample environment, detector tank, and the position-dependent efficiency function of the detector. Vanadium measurements are typically done in conjunction with sample measurements and serve as a normalisation basis. 

For the present Multi-Grid measurements, a vanadium rod with a 6.35~mm diameter and 50~mm length was used. These measurements have allowed us to compare very similar scattering intensities detected by the $^3$He detectors and the Multi-Grid. It should be noted that the neutron flux need not be identical, due to possible variation in other parts of the instrument. However, as a first approximation we can consider that a very similar flux of elastically scattered neutrons are seen by both detectors. 

\begin{figure}[tbp] 
\centering
\includegraphics[width=0.99\textwidth]{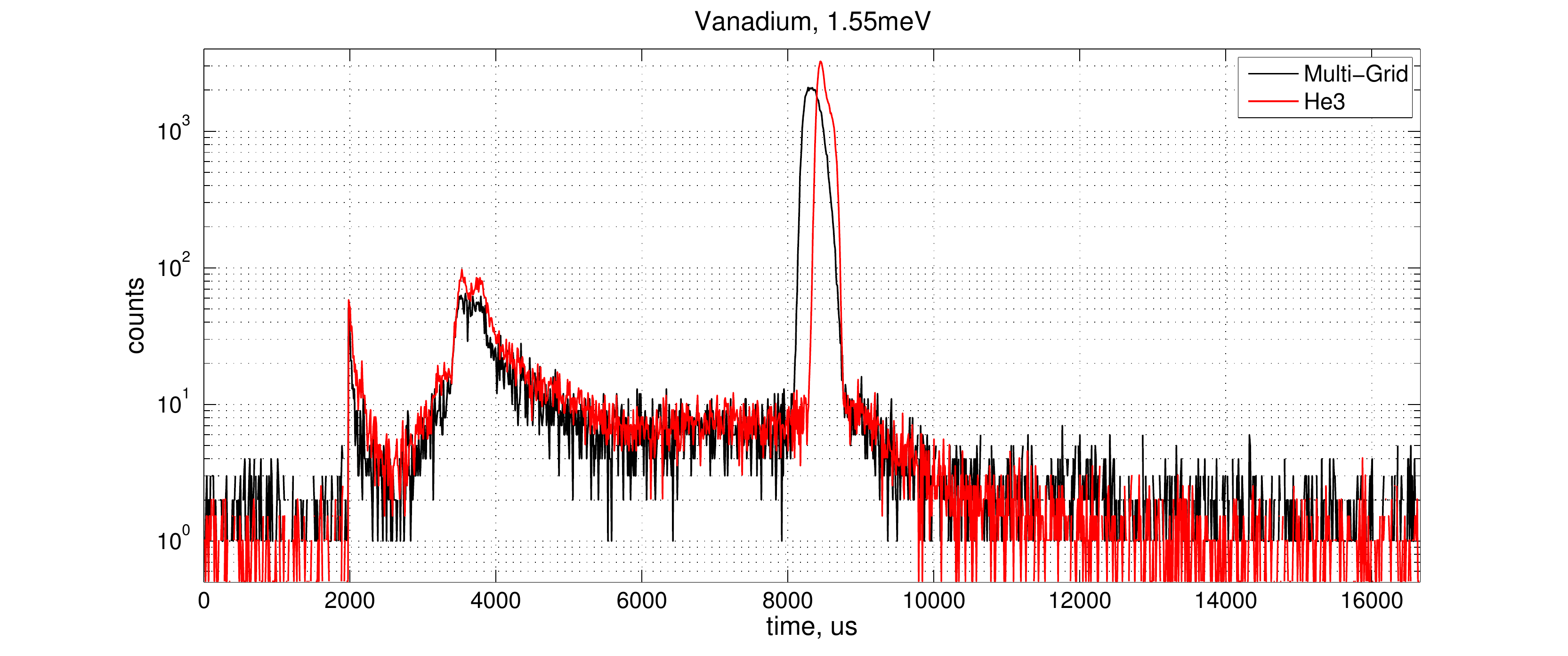}
\caption{Time-of-Flight spectrum of vanadium sample measured with Multi-Grid and $^3$He detectors for incident neutron energy of 1.55~meV. The raw detection time is plotted, no information about the geometry of the detectors is used here. Several features are visible: the prompt pulse (2000~$\mu$s), neutron energy gain (3000-5000~$\mu$s), elastic peak (8500~$\mu$s). The full time scale corresponds to the period of the SNS pulse, 16667~$\mu$s. The time axis is offset so that the elastic peak is at the middle of the range. The time offset with respect to the beam-on-target time is known for each beam setting.}
\label{fig:tofspec}
\end{figure}

An example of a ToF spectrum measured with a vanadium sample is shown in figure~\ref{fig:tofspec}, where the main features can be seen. This spectrum is plotted for the time equal to the period of the source pulse and is accumulated over many pulse periods. The relative time of the elastic peak, the inelastic region and the prompt pulse will vary depending on the incoming energy, since their arrival time depends on the neutron energy.

\subsubsection{Energy Transfer Determination}

The final energy, $E_f$, is determined based on the ToF of the neutron, combined with the knowledge of the initial energy, $E_i$. The beam delivery system selects a narrow band of $E_i$ using a series of choppers. This system works purely by the selection of neutrons from the primary pulse, which contains neutrons spread over a wide energy range. Therefore, each neutron arriving at the sample can be considered to have travelled directly from the moderator. The information on its expected energy is given by the setting of the chopper cascade. Neutron energies in $meV$ can be expressed as velocities in $m / s$ using, 
\begin{equation}
v_i = \sqrt{191314 \times E_i},  v_f = \sqrt{191314 \times E_f}
\end{equation}
The arrival time at sample, in $s$, is 
\begin{equation}
t_s = L_1 / v_i + T_0
\end{equation}
where $L_1$ is the distance from source to sample. The final velocity for a neutron detected in the voxel $x$ at the time $t_{ToF}$ is then
\begin{equation}
v_f = {L_x} / (t_{ToF}-t_s)
\end{equation}
where $L_x$ is a distance from the sample to the voxel $x$. In the above, $T_0$ is the expected delay for a neutron of energy $E_i$ to be emitted from the moderator after the proton pulse impact on the target. This delay generally increases for lower energies. Finally, we define energy transfer as $E = E_i-E_f$, which is positive when the neutron transfers part of its energy to the sample. 

\begin{figure}[tbp] 
\centering
\includegraphics[width=0.49\textwidth]{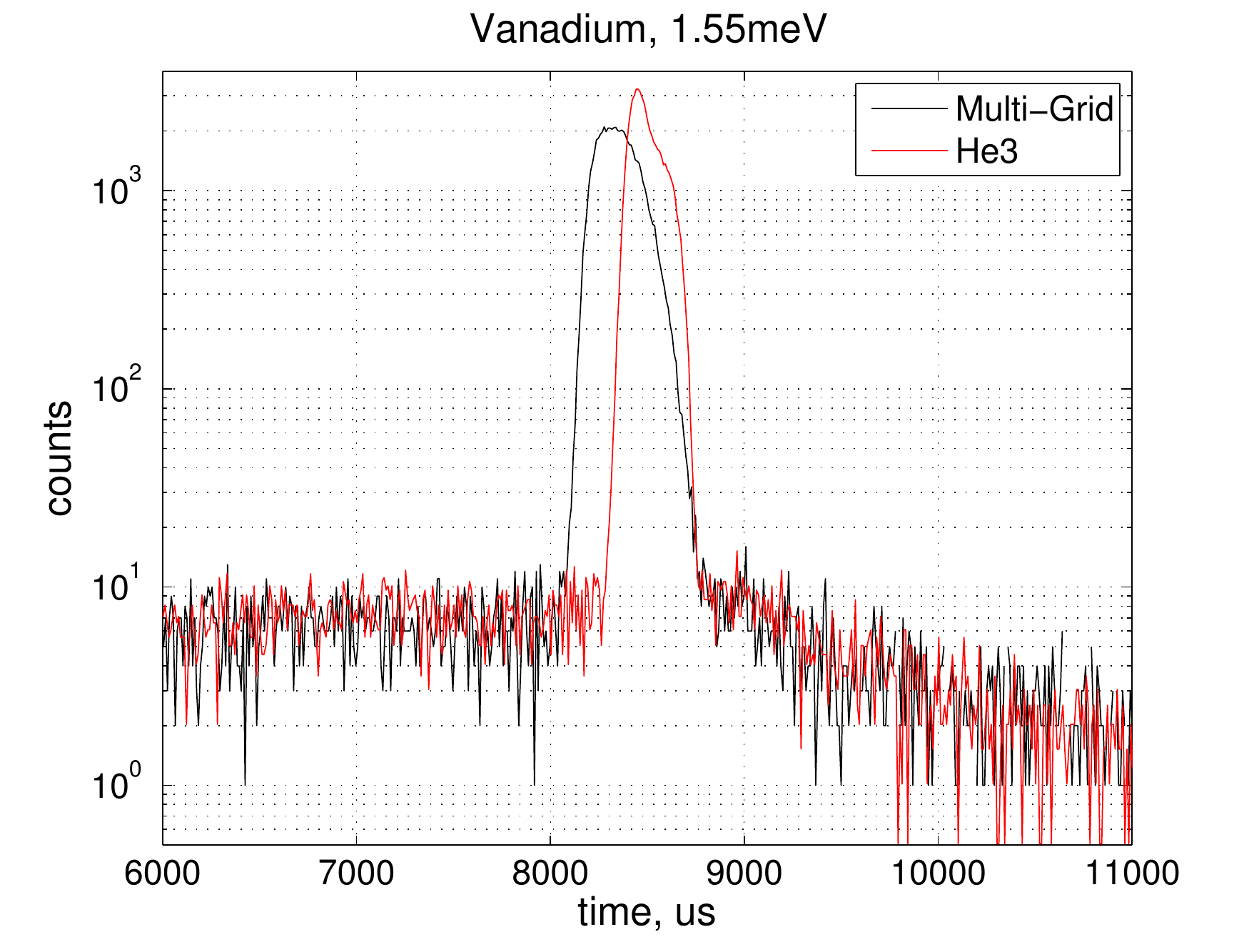}
\includegraphics[width=0.49\textwidth]{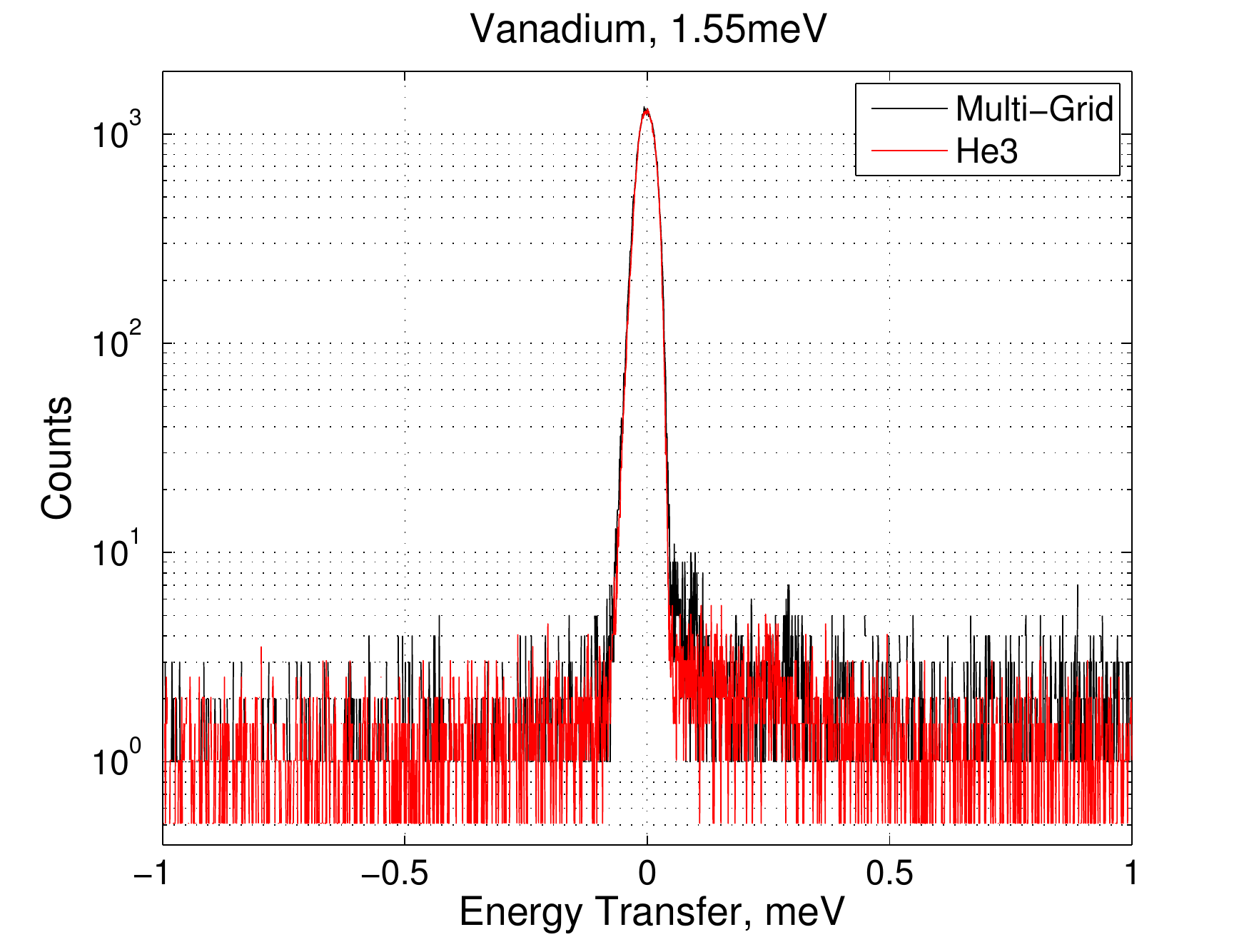}
\caption{ToF to energy transformation for a range around the elastic line. \textbf{Left:} the raw ToF spectra of Multi-Grid and $^3$He detectors. \textbf{Right:} the resulting energy transfer spectrum.}
\label{fig:etransform}
\end{figure}

\begin{figure}[tbp] 
\centering
\includegraphics[width=0.70\textwidth]{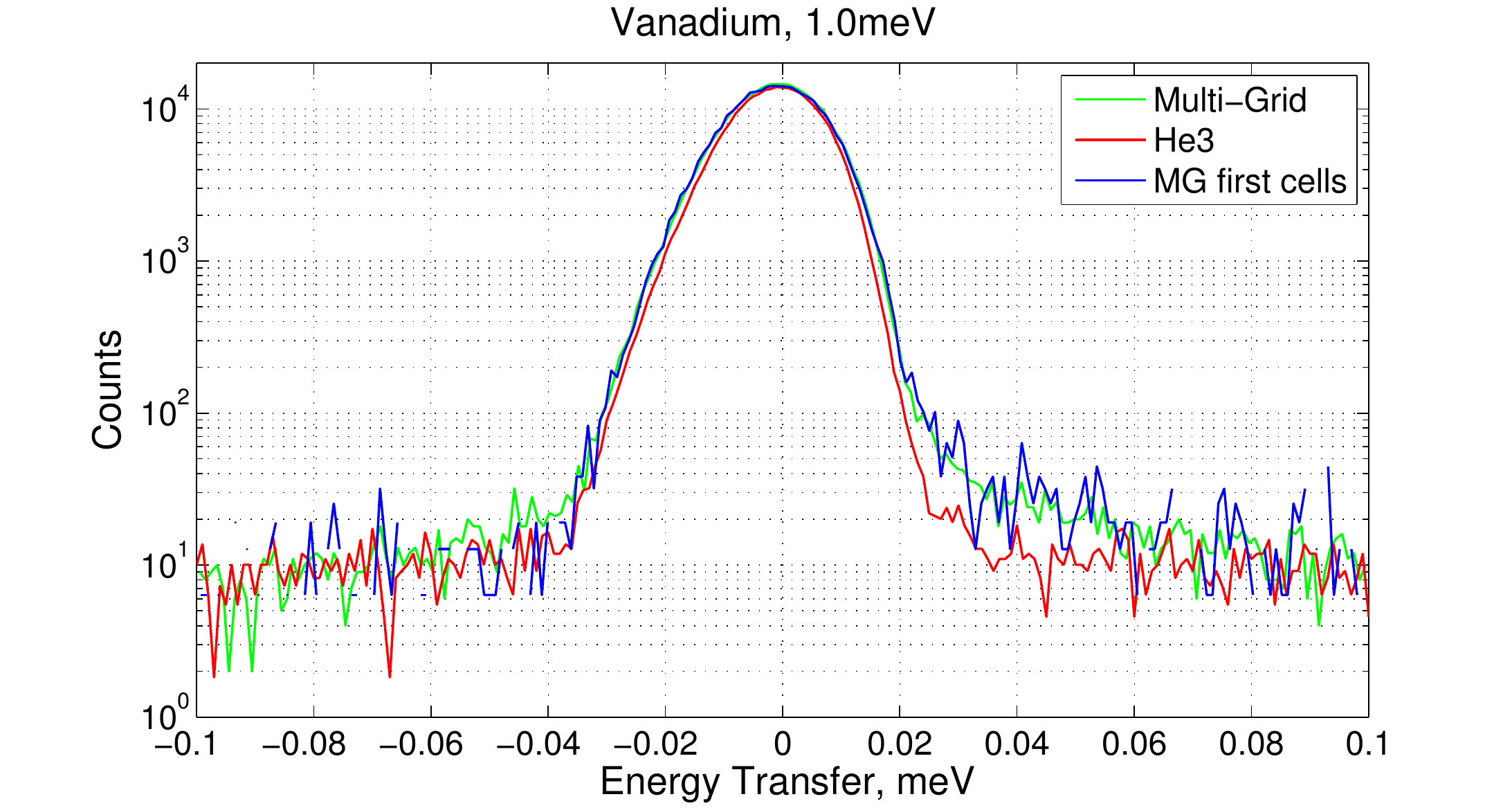}
\caption{Energy transfer reconstruction where only the front layer of the detector is used compared to reconstruction based on the full detector.}
\label{fig:etransform2}
\end{figure}

The transformation above was performed for both Multi-Grid and $^3$He detector data. An example is shown in figure~\ref{fig:etransform}, where both the raw ToF spectrum and the resulting $E$ spectrum are shown. While Multi-Grid has a segmentation in depth of the detector, this does not influence the precision of the $E$ reconstruction, as position of interaction is known to within the size of a voxel and the position of each individual voxel and its distance to the sample, $L_x$, is known. This is further demonstrated in figure~\ref{fig:etransform2}, where the reconstruction for the Multi-Grid is performed taking only the single front-most layer of voxels. The resulting distribution is the same as using the full depth of the detector (the statistics are, of course, lower, therefore, here, the front-voxel curve has been normalised to match the others and shows a greater level of noise.).

\subsubsection{Elastic Energy Resolution}

The FWHM of the elastic peak in the energy scale is referred to as \emph{elastic energy resolution} of the instrument and is one of its main specifications. This quantity is a function of the settings of the instrument and describes the energy resolution of the entire instrument as a whole. Several resolutions combine to form it: the energy spread and time spread of the beam pulses due to the primary spectrometer, size of the sample, distance used in ToF measurement and the time resolution of the detector. In this section, we look at the elastic energy resolution obtained simultaneously with the Multi-Grid and $^3$He detectors as a function of $E_i$. Note that the detector is only one of several factors influencing the energy resolution, however, since the data comes from the same measurements, the other factors are constant. We define the energy resolution as the FWHM of the elastic peak in the energy scale. 

The instrument can be used with a variety of settings depending on the needs of experiments being performed. As common for many instruments, CNCS can be tuned to increase flux at a cost of the energy resolution or vice versa. The settings are mainly defined by the chopper phasing, opening times and rotation speeds~\cite{cite:cncs}. Additionally, the size of the sample introduces an uncertainty in the distance between the scattering and detection locations, and thus will also affect energy resolution. 

\begin{figure}[tbp] 
\centering
\includegraphics[width=0.95\textwidth]{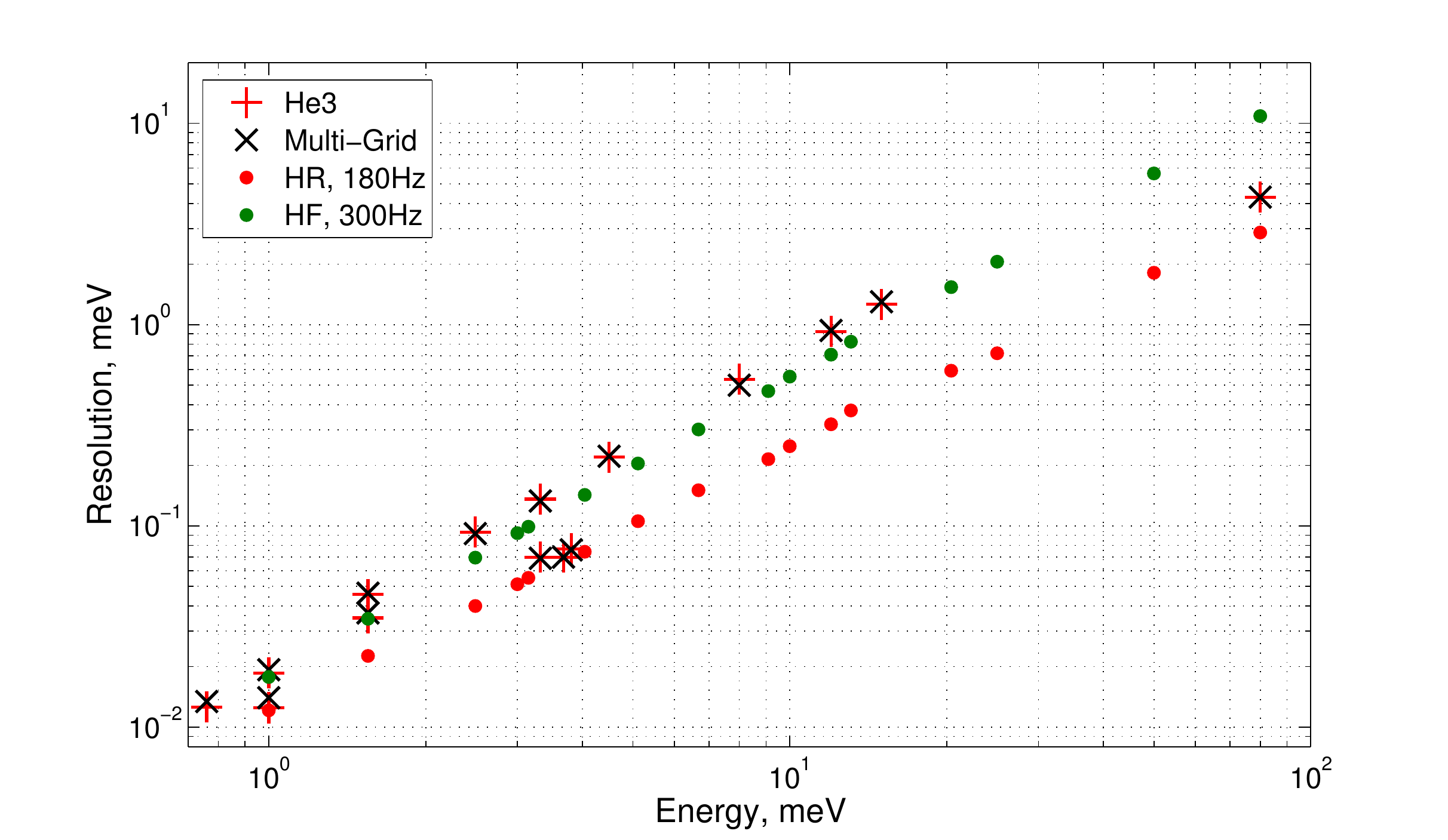}
\caption{Instrument elastic energy resolution measured with the Multi-Grid and $^3$He detectors. Data are compared and overlaid onto published CNCS data~\cite{cite:cncs} for High Resolution (HR) and High Flux (HF) settings from earlier measurements.}
\label{fig:eres}
\end{figure}

For the settings of the instrument used to obtain the measured range of incoming neutron energies, the width of the elastic line decreases with decreasing energy. Table~\ref{tab:eres} and figure~\ref{fig:eres} show this dependence. The energies that have been measured range from 0.76~meV to 80~meV, and the corresponding resolutions range from 0.0134~meV to 4.3~meV, or in relative terms 1.4\% to 5.37\%. These values are essentially equivalent to those extracted from the $^3$He detector data, demonstrating that both detector technologies perform equivalently. Note that there were 4 sets of measurements performed. These are denoted 1-4 in the table. Measurements 1 and 2 are at a high flux setting and 3 and 4 are in high resolution setting. We can see that each setting results in values lying on a straight line in this log-log plot. These lines are different between all 4 measurements, indicating that the energy resolution is sensitive to the exact settings of the instrument.

\begin{table}[tbp]
\caption{Elastic energy resolution of the instrument as a function of neutron energy measured based on data collected with the two detectors. Measurements 1 and 2 are at the high flux setting and 3 and 4 at the high resolution setting.}
\label{tab:eres}
\smallskip
\begin{tabular}{llllr}
\hline
Energy, meV 	& Wavelength, \AA 	& dE in MG, meV	& dE in $^3$He, meV  & MG resolution, \% \\ 
\hline
Measurement 1 \\
0.76		&	10.4		&	0.0134		&	0.0126		&	1.76\\
1.0		&	9.04		&	0.0193		&	0.0186 		&	1.93\\
1.55		&	7.20		&	0.0370		&	0.0349		&	2.39\\
\hline
Measurement 2 \\
1.55		& 	7.20		&	0.0463		& 	0.0457		&	2.99\\
2.49		&	5.73		&	0.0915		&	0.0933		&	3.67\\
3.32		&	4.96		&	0.133		&	0.136 		&	4.01\\
4.5		&	4.26		&	0.222		&	0.220 		&	4.93\\
8		&	3.20		&	0.500		&	0.535 		&	6.25\\
12		&	2.61		&	0.935		&	0.925 		&	7.79\\
15		&	2.34		&	1.30			&	1.26 			&	8.67\\
\hline
Measurement 3 \\
3.32		&	4.96		&	0.069		&	0.070 		&	2.08\\
\hline
Measurement 4 \\
1.0		&	9.04		&	0.0140		&	0.0125 		&	1.4\\
3.678	&	4.716	&	0.070		&	0.070 		&	1.9\\
3.807	&	4.635	&	0.076		&	0.077 		&	2.0\\
80.00	&	1.01		&	4.30			&	4.30	 		&	5.37\\
\hline
\end{tabular}
\end{table}

\subsubsection{Relative Efficiency}
\label{sect:eff}

\begin{figure}[tbp] 
\centering
\includegraphics[width=0.90\textwidth]{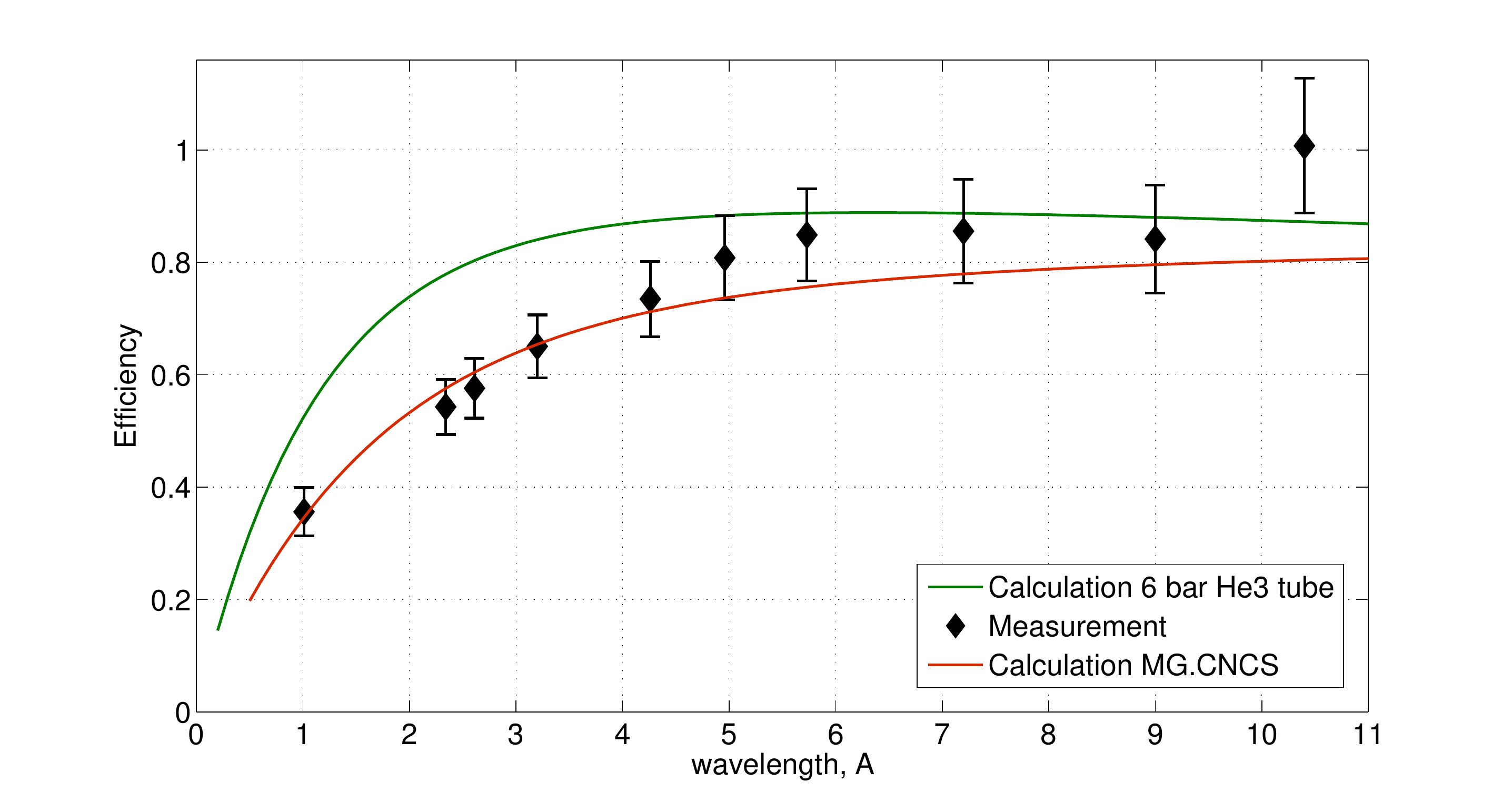}
\caption{The calculated wavelength-dependent efficiency of the MG.CNCS and 6 bar $^3$He tubes. The data points show measured efficiency of MG.CNCS obtained by multiplying the relative efficiency in table~\ref{tab:releff} by the calculated efficiency of a 6-bar $^3$He tube.}
\label{fig:releff}
\end{figure}

Efficiency is another key specification of a detector. In order to accurately measure efficiency, one generally needs to use a well defined beam that does not change in time to measure the counts in the detector being characterised. The beam flux can be calibrated using a reference detector~\cite{cite:piscitellithes}. The ratio of the detected rate to the total rate in the beam then provides the sought efficiency. The test setup at CNCS is not ideally suited for a high precision efficiency measurement, since a reference detector is not available and the detectors are exposed to a flux of neutrons with possible angular non-uniformities and with a spread in energy, rather than a well defined beam. We can, however, make comparisons between the rate in $^3$He and Multi-Grid detectors in measurements with a vanadium sample where the flux is expected to be approximately uniform in all directions. This does not guarantee that the measurement is as accurate as a test beam measurement, as there may be differences in the flux measured by the two detectors. The comparison between the counts measured by Multi-Grid and $^3$He detectors in the vanadium measurements available is still a good estimate of the efficiency of the Multi-Grid detector and is presented in table~\ref{tab:releff} and visualised in figure~\ref{fig:releff}. The counts in the elastic peaks in the energy scale were integrated for each energy to obtain these results. 

In order to estimate the systematic uncertainty of the relative efficiency, we need to have an estimate of the expected difference in flux between $^3$He bank \# 50 and the Multi-Grid. To do this, we considered the variation of the elastic peak counts from each of the 50 banks of $^3$He detectors to its closest neighbour. The MG detector can be seen as the nearest neighbour of bank \# 50. The mean difference between neighbour counts is then treated as an estimator of the deviation between the the flux in the MG and in bank \# 50. The $^3$He banks where the count rate deviated significantly from the rest of the array due to the proximity of the transmitted beam were excluded from this calculation. Note that the statistical uncertainty is much smaller than this variation, less than 1\% for every measurement, and therefore was neglected. Comparing the counts in the MG to that of the 6-bar $^3$He tubes, the overall trend is as expected -- the MG is very close in efficiency at long wavelengths and falls towards 70\% of $^3$He tube approaching the shorter wavelengths. The measurement is consistent with calculation, with the exception of the point at the longest wavelength, where the MG counts are significantly above expectation. 

\begin{table}[tbp]
\caption{List of available vanadium measurements with information on relative efficiency between Multi-Grid and $^3$He detectors. Note that only total counts are shown, not taking into account the durations of the measurements or the incoming flux.}
\label{tab:releff}
\smallskip
\begin{tabular}{llllr}
\hline
Energy, meV  	& Wavelength, \AA 	& Counts in MG	& Counts in $^3$He		& ratio MG/$^3$He \\ 
\hline
0.76		&	10.4		&	262293		&	227090		&	1.15 $\pm$ 0.12\\
1.00		&	9.0		&	250237		&	261780		&	0.956 $\pm$ 0.096\\
1.55		&	7.2		&	349035		&	385973		&	0.964 $\pm$ 0.092\\
2.49		&	5.73		&	164505		&	172160		&	0.956 $\pm$ 0.082 \\
3.32		&	4.96		&	258246		&	282220		&	0.915 $\pm$ 0.075 \\
4.5		&	4.26		&	195881		&	232870		&	0.841 $\pm$ 0.067 \\
8		&	3.20		&	235545		&	304250		&	0.774 $\pm$ 0.056 \\
12		&	2.61		&	276897		&	386390		&	0.717 $\pm$ 0.053 \\
15		&	2.34		&	218089		&	313120		&	0.697 $\pm$ 0.049 \\
80		&	1.01		&	20519		&	30315		&	0.677 $\pm$ 0.043 \\
\hline
\end{tabular}
\end{table}

\subsubsection{Aluminium Bragg Edge}

A set of vanadium sample measurements were performed at several energies very close to the Bragg edge of aluminium, at 3.742~meV, or $\lambda$=4.676~\AA. Bragg reflection is described by the following formula,
\begin{equation}
n \lambda = 2 d \sin (\theta /2)
\end{equation}
where $\lambda$ is the wavelength, $n$ is an integer, $d$ the lattice spacing and $\theta$, the scattering angle. For aluminium, $\theta=180^\circ$ (\emph{i.e.} backscattering) corresponds to $\lambda=4.676$~\AA ~and smaller angles for shorter wavelength. For longer wavelengths, Bragg's law cannot be satisfied, and therefore, no coherent reflection occurs and only the incoherent scattering cross section is active. Due to this reason, we expect the losses due to scattering in Al to change sharply at this energy. Scattered neutrons travel backwards through the detector and can be detected after a longer time than they normally would for that detection position. The energy reconstruction based on ToF will see them as slower, or lower energy than they really are. We therefore expect that detection of scattered neutrons will appear in measurements on the neutron energy loss side of the elastic peak. This effect is expected and was first observed in the measurement on the IN6 demonstrator~\cite{cite:in6test}. On CNCS, we were able to investigate it in more detail, since the energy of incoming neutrons can be freely selected on CNCS, including energies immediately adjacent to the Bragg edge. 

\begin{figure}[tbp] 
\centering
\includegraphics[width=0.49\textwidth]{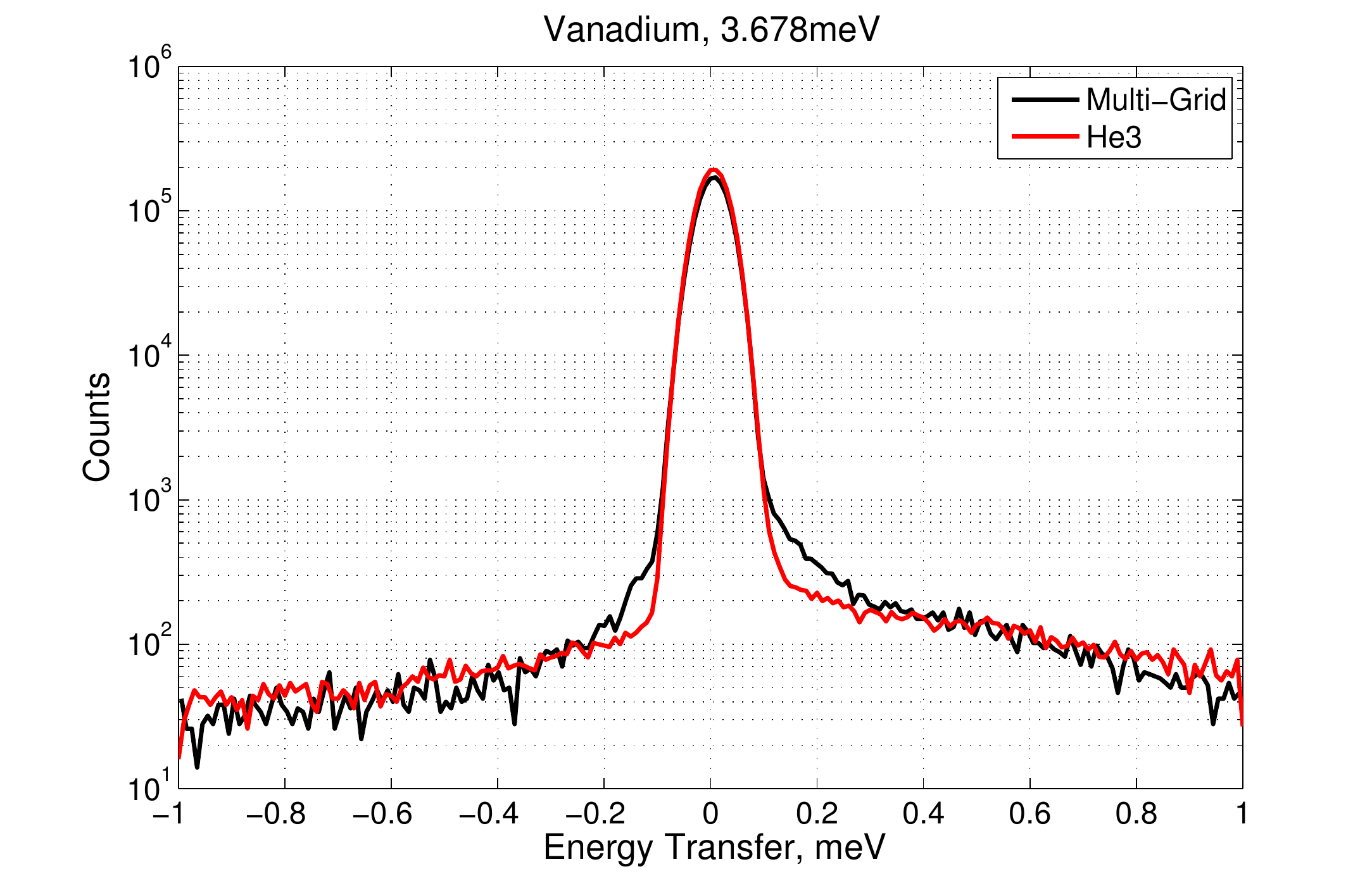}
\includegraphics[width=0.49\textwidth]{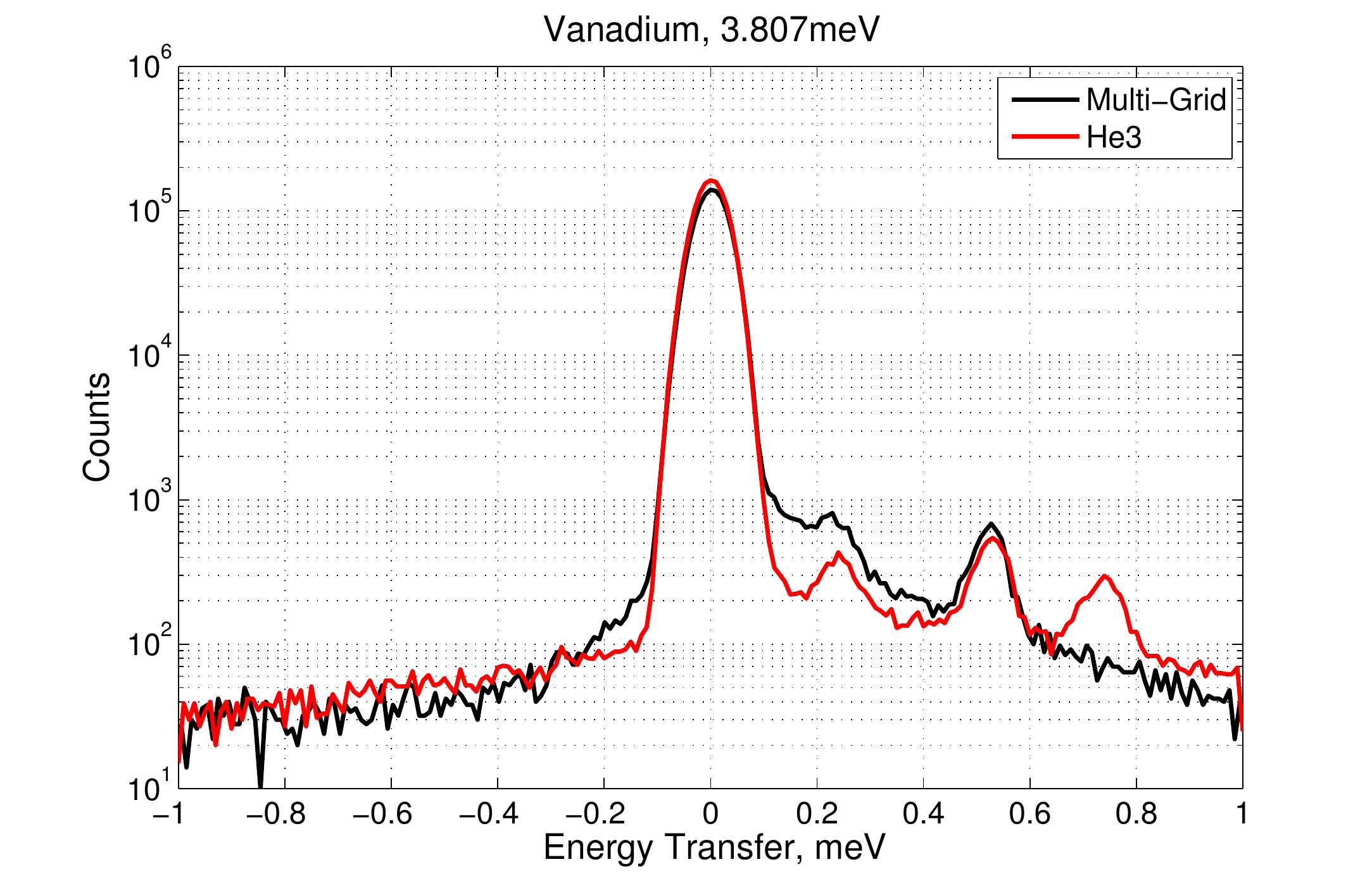}
\caption{Comparison of the elastic line above and below the Bragg edge energy. Measurements of Multi-Grid and $^3$He at 3.678~meV (left) and 3.807~meV (right) are shown. These energies were chosen so that, at the energy resolution in this setting, the incoming energy distributions are fully above or below the Bragg edge.}
\label{fig:braggedge}
\end{figure}

In addition to the coherent scattering in the detector itself, one has to keep other materials traversed by neutrons in mind. For instance, if there is a cylindrical aluminium cryostat enclosing the sample, neutrons scattered from the sample in the direction opposite to the detector, can then reflect from the cryostat and then reach the detector. These will have travelled an extra distance equal to the diameter of the cryostat. A similar effect can occur due to the Al window between the sample environment and the Ar-filled detector tank. In case of scattering within the detector, the most common increase in the flight distance would be 2~cm, \emph{i.e.} the extra distance travelled by a neutron when it has scattered in a voxel and was detected in the preceding voxel on its way back to the front of the detector. Figure~\ref{fig:braggedge} shows comparisons of the elastic lines for energies just below and just above the Al Bragg edge cut-off. There are several features that we see appearing when the reflection is active. There is a shoulder at the right of the peak -- this is short-range scattering, consistent with scattering in the detector. Indeed we see that this is slightly enhanced in the case of the Multi-Grid detector compared to $^3$He detectors. Note that the contribution of the most common scattering case in the detector (that resulting in 2~cm extra flight distance) cannot be resolved in this plot, as the resulting energy would be inside the elastic peak. The apparent increase on the right-hand side of the peak corresponds to 4-5 extra boron layers traversed (or 8-10~cm of additional flight distance). There are also several peaks further away from the elastic line. These are too separated from the line and are not consistent with in-detector scattering. Indeed, they appear in both Multi-Grid and $^3$He detector data in equal measure. We therefore interpret this as the effect of the sample environment. Surprisingly, there is one peak that only occurs in the $^3$He detectors. 

This measurement shows that the shoulder due to Bragg scatting is more significant in the Multi-Grid detector. However, it is clear, that ideally, measurements should avoid this incoming energy altogether, since reflections from all Al components in the beam path will create artefacts in the data. It should be noted that these effects can be simulated and this will be presented in a separate publication where the GEANT4-based framework~\cite{cite:dgframework, cite:dgframework2} is used to compare measured data to simulation.

\subsection{Single Crystal Sample}

While a vanadium sample scatters neutrons incoherently, the opposite is true in single crystal measurements. A crystal presents a continuous lattice with a constant orientation and d-spacing, making it possible for neutrons to reflect coherently, and hit the detector in a spot. This presents the most extreme situation for the detector as neutrons are both focused in time, as well as in position, resulting in a very high local instantaneous rate~\cite{cite:stefanescu}.

An intermediate case is diffraction from a polycrystalline powder. A powder can be seen as a collection of small crystals that are randomly oriented, but all have the same d-spacing. This means that the coherent reflections still have a fixed angle with respect to the beam, but are randomly oriented azimuthally, resulting in a cone-shaped scattering pattern. When incident on a cylindrical detector, the rings (arcs) that are seen are called Debye-Scherrer cones. In this case, the intensity is still focused in time, but is spread over a number of pixels, which can be in the order of 100 pixels or more. 

We performed a measurement with a single crystal of UGe$_2$. Two settings of the crystal rotation and incoming wavelength were chosen so that one of the Bragg reflections was directed to the Multi-Grid detector in one case and to the edge of the $^3$He array in the other case. In this way, a very similar high-intensity feature could be studied in both detectors. An incident neutron energy of 13.74~meV was used to target the MG and 17.20~meV was used to target He3.

The second interesting feature of UGe$_2$ is due to its U content. The crystal was grown using depleted uranium. Nevertheless, the traces of the fissile isotope, $^{235}$U, are sufficient to see some fission reactions when the sample is irradiated with neutrons. These neutrons are much more energetic than the cold neutrons delivered by the beam, and it is interesting to investigate their effect on the detectors. Two sets of measurements with UGe$_2$ were performed. The first was aimed at investigating the effect of the intense Bragg reflection on the detector response. The second investigated the ToF spectra, including the effect of the fast neutrons from U. These are described in the following sections.

\subsubsection{High Rate in the Bragg Reflection}

The Bragg reflection from the crystal gives us the highest instantaneous rate. It is interesting to compare the performance of the $^{10}$B and $^3$He detectors in these conditions, since for future instruments at ESS, gains of one or more order of magnitude in flux are expected. The rate capability can be limited by the detector itself, as well as by the associated data acquisition system. In a wire proportional chamber, the bulk of the amplitude signal is generated by gas gain in the high electric field region close to the anode wire. From here electrons are quickly collected onto the wire, while the ions take a much longer time to drift to the cathode. Therefore, the signal can be detected quickly, but a residual space charge remains in the gas for a longer time. This charge will reduce the local electric field, thus lowering the gas gain experienced by following detections. The gas gain used in the Multi-Grid detector is approximately a factor 50, while typically about a factor 500 gain is used in $^3$He tubes in order to maximise position resolution via the charge division readout. Additionally, the readout electronics requires a certain time to process each event, and a subsequent event occurring within this dead time, will be missed. Therefore, at a high rate, we can expect a loss of counts due to both of these effects.

In both acquisition systems, that of $^3$He and for the Multi-Grid detectors, an electronics dead time of approximately 1.5~$\mu$s was found. In the case of $^3$He, the dead times only apply to individual tubes, \emph{i.e.} a detection in one tube, does not prevent a detection in another. In the electronics used in this test with the MG, this dead time is global due to the use of the multiplexing front-end. The peak instantaneous rate that was obtained in the most intense of these measurements corresponds to 150~kHz in the spot (about 30 neutrons during an elastic peak in a single time frame). Therefore, the dead time presents only a minor limitation, as it is an order of magnitude smaller than the average time between events in the most intense measurements. 

\begin{figure}[tbp] 
\centering
\includegraphics[width=1.07\textwidth]{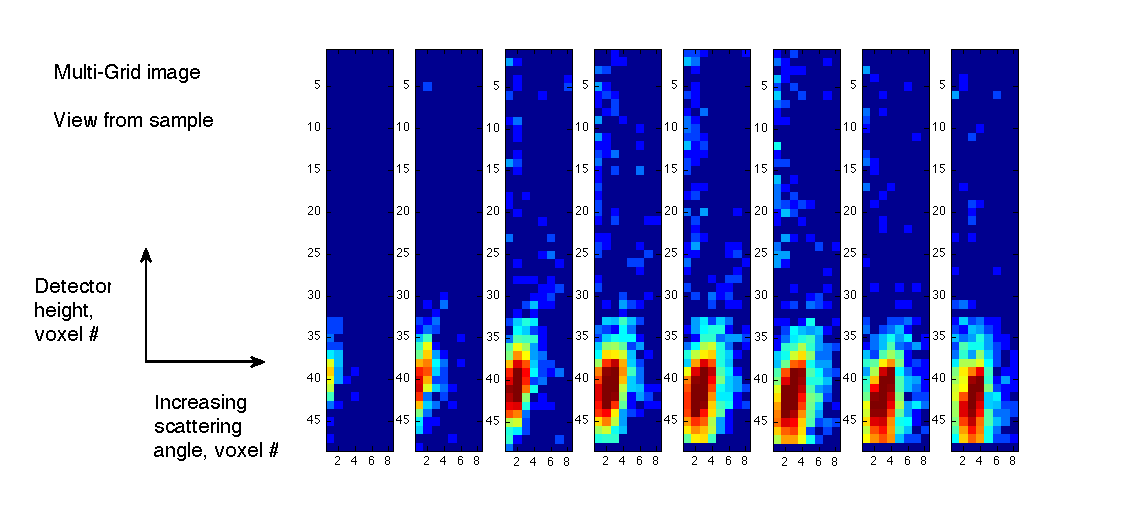}\\
\includegraphics[width=1.07\textwidth]{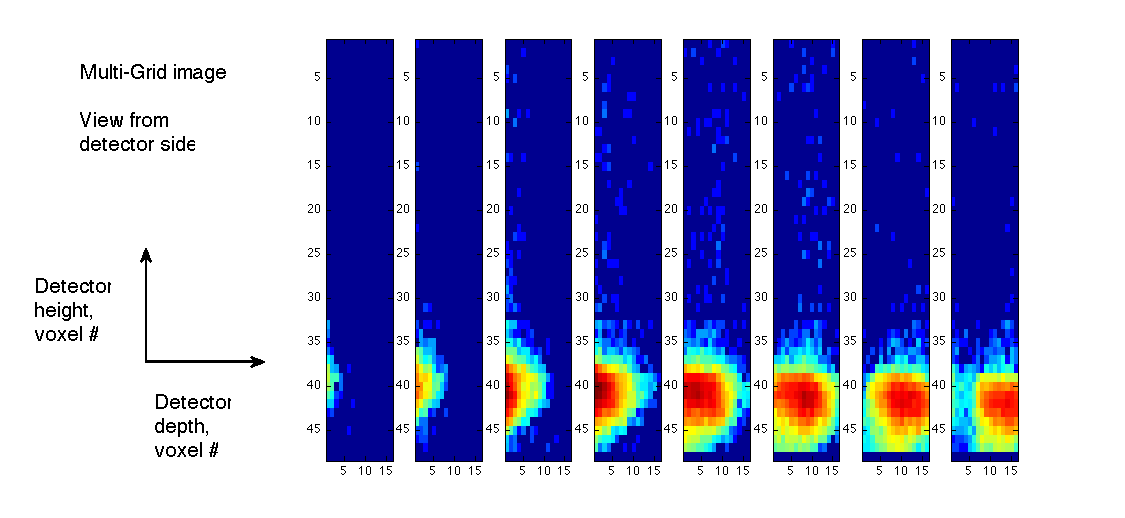}
\caption{Evolution of the Bragg reflection on the surface of the MG detector. The top row of images show the front projection and the bottom row show the side projection of the image in the detector. Each image, from left to right, advances by 30~$\mu $s. The pulse of neutrons can be seen traveling from the front of the detector to the rear. The position of the pulse shifts by approximately 2 pixels to the right and down over the duration of the pulse, as seen in the front projection. This is due to the variation of the incoming energy in the pulse -- and therefore, reflection angle (as far as allowed by crystal mosaicity and beam divergence).}
\label{fig:imageframesmg}
\end{figure}

\begin{figure}[tbp] 
\centering
\includegraphics[width=1.07\textwidth]{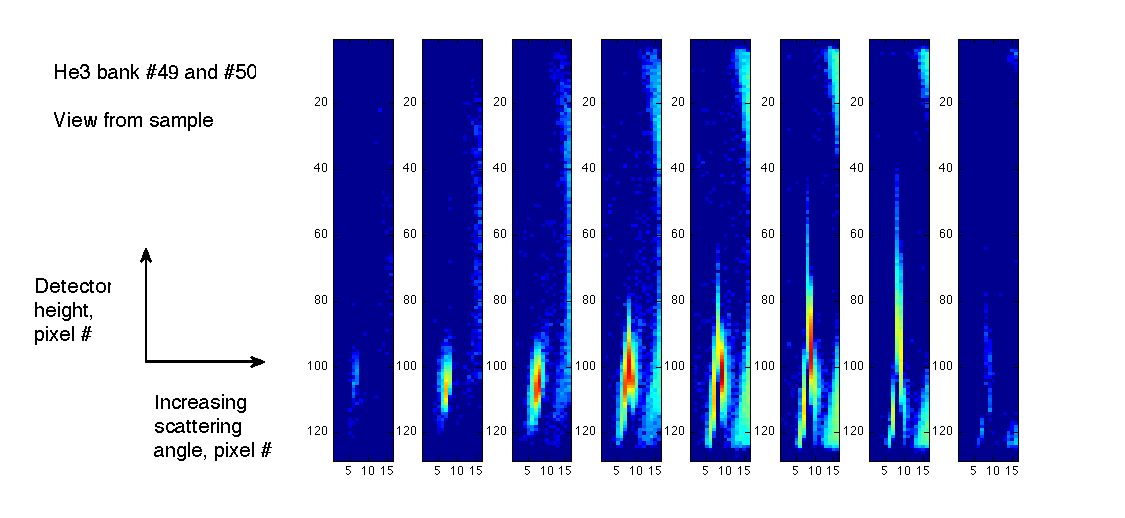}
\caption{Evolution of the Bragg reflection in 16 tubes of the $^3$He detector array. Each image, from left to right, advances by 30~$\mu $s. Large variations of the reconstructed position along the length of the tubes can be seen as a function of time for tubes that experience the highest intensity. The arch-shaped feature that appears in the right top and bottom corners of the images is a powder-like reflection from aluminium in the sample environment. Note that 16 of the 2-m tubes are shown, \emph{i.e.} twice the width and twice the height of the Multi-Grid detector image in figure~\ref{fig:imageframesmg}.}
\label{fig:imageframeshe3}
\end{figure}

A significant effect that we found is a loss of position sensitivity in the $^3$He tubes during the pulse. Figure~\ref{fig:imageframesmg} shows the evolution of the Bragg reflection on the surface of the Multi-Grid with time, and figure~\ref{fig:imageframeshe3} shows the equivalent image for $^3$He array. We see that in both cases it starts as a spot several pixels across. In the Multi-Grid detector we see that it drifts to the right as time progresses, since later in the incident pulse one probes the longer wavelength part of the beam and scattering angle increases.  

In the $^3$He tubes, the spot extends by as much as 1/3 of the tube length towards the middle of the tubes for the tubes where the intensity is highest. Furthermore, the tube at the centre of the spot initially sees a high rate, but later appears to lose sensitivity. We believe that this is caused by the charge pile up in the tubes. This affects the charge-division readout, which generates the position information, as well as the overall ability to handle incoming events. A similar effect is observed in other position sensitive $^3$He tubes, such as on SEQUOIA~\cite{cite:sequoya}. This is not seen in the Multi-Grid detector, as, on one hand, the rate is spread over 16 detection cells for each voxel, and on the other, a much lower gas gain is used, meaning that the residual space charge from each interaction is lower. Additionally, the position is reconstructed using coincidence and is only sensitive to the simultaneous detection of two signals, whereas in charge division, position reconstruction relies on the comparison of amplitudes readout at two ends of a tube. 

\subsubsection{Fast Neutrons}

For the fast neutron measurement, we varied the energy of the incoming beam and observed the features of the ToF spectra. While there may be Bragg reflections at these settings, they were not hitting either the Multi-Grid detector, nor bank 50 of the $^3$He array. We look instead at neutrons scattered incoherently and fast neutrons, which are emitted isotropically. 

\begin{figure}[tbp] 
\centering
\includegraphics[width=0.90\textwidth]{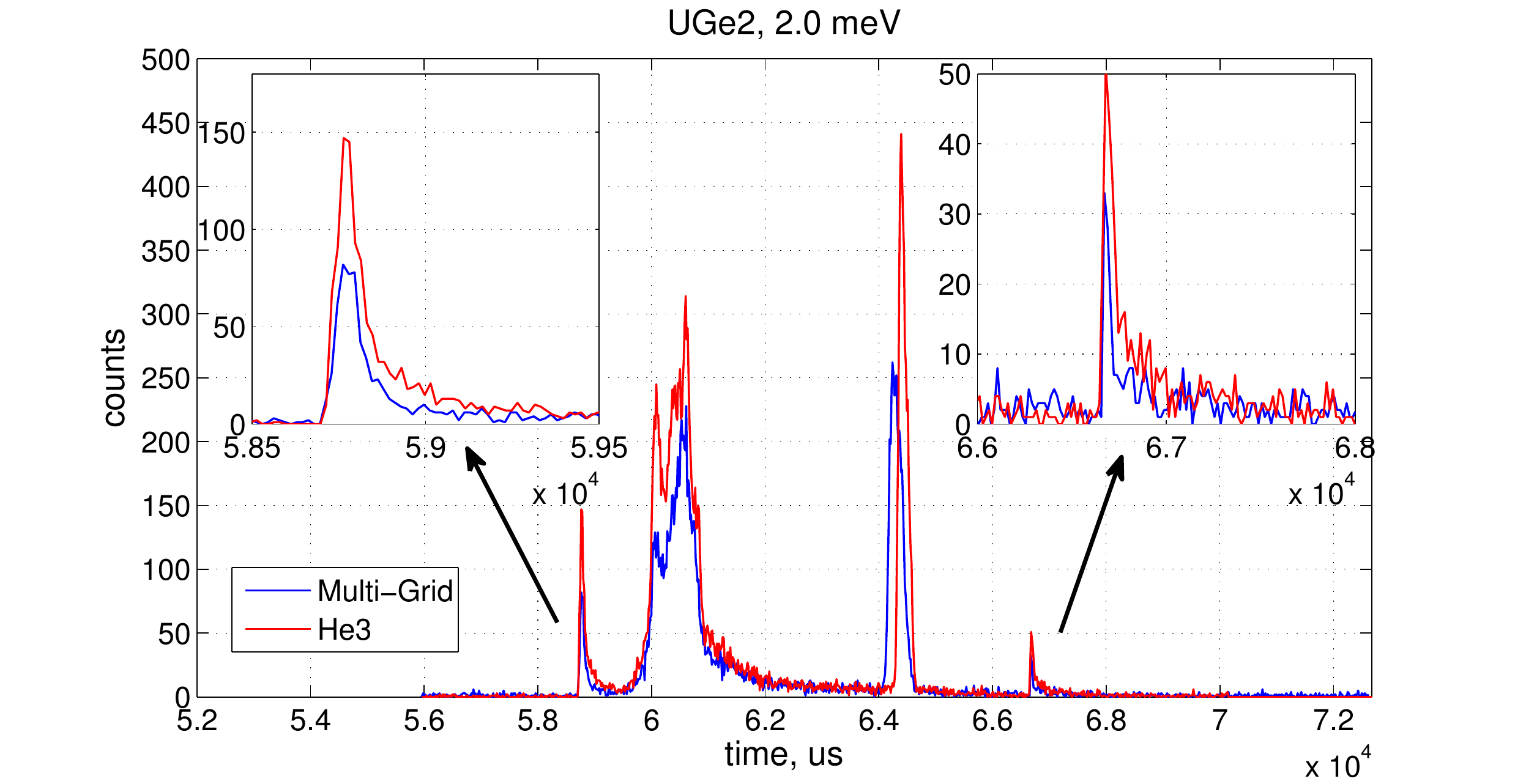}
\caption{ToF spectrum from the UGe$_2$ measurement at 2.0~meV. The inserts show the prompt fast neutron peak from the source at 67000~$\mu$s and the prompt pulse from the sample at 59000~$\mu$s.}
\label{fig:fastn}
\end{figure}

Fast neutrons behave quite differently to thermal neutrons and generally constitute a background in neutron scattering experiments. There is a component of fast neutrons produced in the atmosphere by cosmic rays. These are of course, uncorrelated in time with the pulses of thermal neutrons delivered to the sample, and so contribute to a time-independent background. At a spallation source, fast neutrons are produced by the proton beam collisions with the target. The intention is to moderate as many of these as possible, in order to use them for scattering, however, some escape the spallation target without losing energy in the moderator. Due to their generally low interaction cross sections, these may reach the detectors, creating background. Fast neutrons can also be produced anywhere in the vicinity of the source by high-energy $\gamma$-rays produced in the target via photo-nuclear reactions~\cite{cite:snsbackground}. These processes involve neutrons at energies of a few MeV and higher. The neutrons due to fissions in the UGe$_2$ sample will have a fission neutron spectrum, which is centred around 1~MeV. Compared to the time scales of thermal neutrons, neutrons from the above processes can be considered to travel instantly. Therefore, if detected, they appear as a sharp peak in the ToF spectrum. Figure~\ref{fig:fastn} shows a spectrum from the UGe$_2$ measurement where both the prompt peak from the target, as well as from the sample are visible. 

We find that both the prompt peaks have a fast rise and a slower tail. This can be understood since the beginning of the peak corresponds to the time of creation of fast neutrons (as the travel time is negligible). Some of the fast neutrons will scatter in the chamber losing part of the energy, possibly thermalising. These constitute the tail of the peak. If we compare the heights of the peaks in their beginning, we essentially compare the sensitivities of the detectors to fast neutrons. The same caveat applies as when comparing efficiency in section~\ref{sect:eff} -- the two detectors are not in the same positions and do not necessarily measure the same flux. Nevertheless, we still notice that the prompt peak ratios between Multi-Grid and $^3$He detector spectra are approximately 0.5 for all of the 6 measurements, as shown in table~\ref{tab:fastn}. 

\begin{table}[tbp]
\caption{Measurements with the UGe$_2$ sample.}
\label{tab:fastn}
\smallskip
\centering
\begin{tabular}{lcccc}
\hline
\hline
Incoming energy	&	Total ratio		&	Elastic ratio	&	Inelastic ratio	&	U peak ratio \\
\hline
1.3 meV			&	0.814		&	1.002		&	0.694		&	0.544 \\
2.0 meV			&	0.761		&	0.920		&	0.675		&	0.567 \\
2.5 meV			&	0.769		&	0.928		&	0.693		&	0.494 \\
3.5 meV			&	0.764		&	0.981		&	0.673		&	0.520 \\
8 meV			&	0.756		&	0.784		&	0.703		&	0.464 \\
32 meV			&	0.538		&	0.383		&	0.657		&	0.676 \\		
\hline
\hline
\end{tabular}
\end{table}

\begin{figure}[tbp] 
\centering
\includegraphics[width=0.65\textwidth]{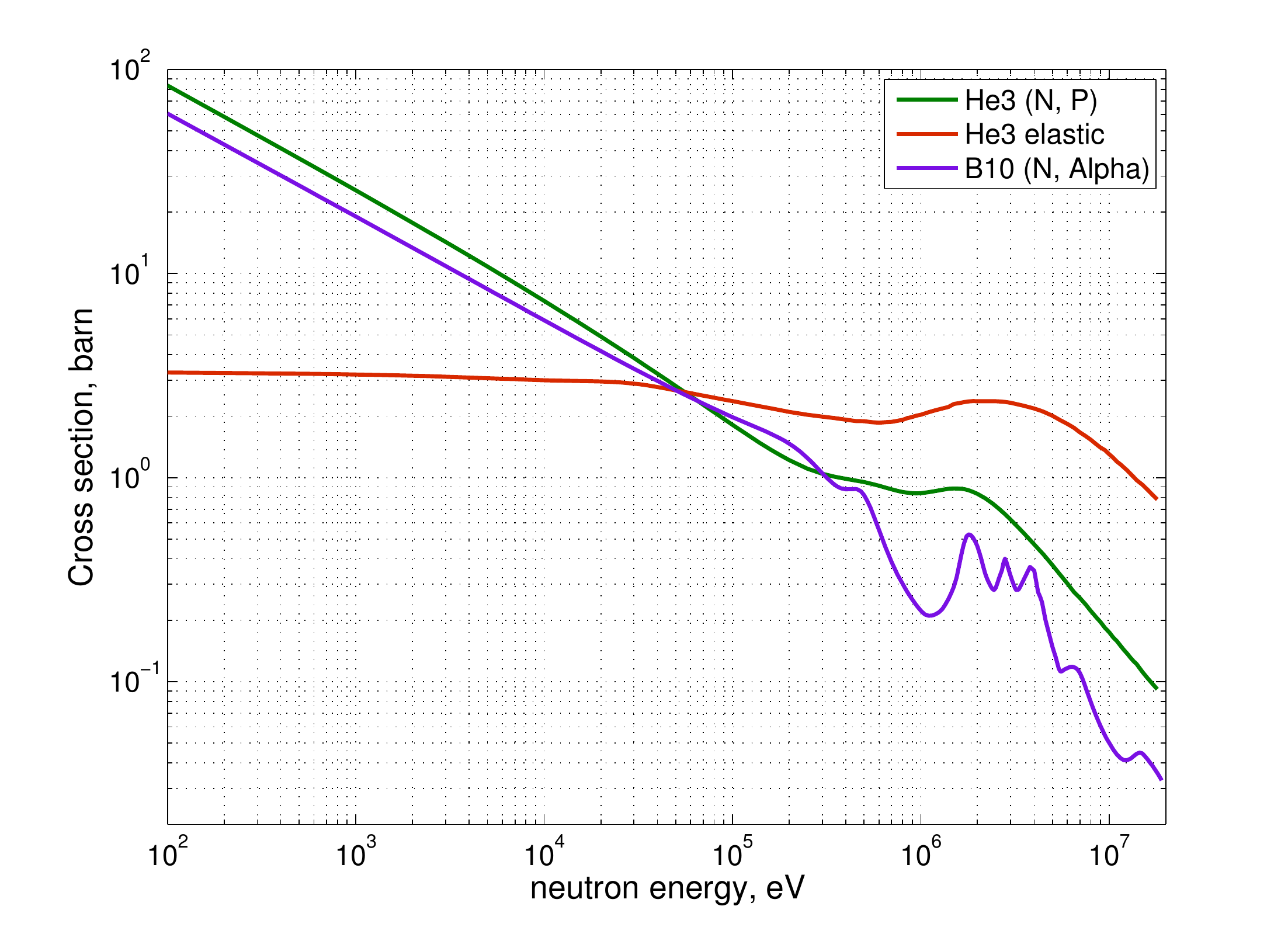}
\caption{Cross sections that contribute to fast neutron sensitivity in $^3$He and $^{10}$B detectors at high energy~\cite{cite:sigma}. Note that in the nomenclature used here, \emph{elastic} cross section refers to all scattering not involving nuclear reactions.}
\label{fig:fastxsect}
\end{figure}

In order to understand the apparent factor 2 increased sensitivity of the $^3$He detectors to fast neutrons, we need to investigate the cross sections involved, see figure~\ref{fig:fastxsect}. The same conversion cross sections of $^3$He and $^{10}$B that allow us to detect thermal neutrons, will also detect fast neutrons, although these cross sections are far lower for high energies. Other nuclear reactions, for example (n, p) or (n, $\alpha$) on other isotopes become possible only from $\approx$10~MeV onwards, and therefore should not be relevant for the 1~MeV fission neutrons. Other types of reactions include the inelastic scattering on nuclei -- the same reaction that allow us to moderate neutrons. In this case, a neutron will transfer a portion of its energy to a nucleus, creating a fast ion. The ions created inside solids are unlikely to escape into the gas (most atoms inside the solid parts of our detectors are heavier than, for example, $\alpha$ produced in the $^{10}$B conversions, and therefore will have shorter ranges). We therefore look at the gases. In helium tubes, the $^3$He itself is the most common gas. It is also a light gas, meaning that more energy can be transferred to it by a neutron in a single collision. In case of the Multi-Grid detector, argon contributes most, however, it is 40 times heavier than a neutron, meaning that the energy transferred is far lower than 1~MeV and likely to be under the detection threshold, additionally the scattering cross section is lower than for $^3$He. We therefore expect that overall, the fast neutron sensitivity should be higher in helium tubes due to the extra inelastic scattering interactions on the $^3$He nuclei\footnote{The same mechanism, energy loss by inelastic scattering, is used in $^4$He detectors for fast neutrons\cite{cite:jebali}.}. This qualitatively agrees with our measurements. 

\subsection{Background}

The signal-to-noise ratio is a key parameter of a chopper spectrometer. As can be seen in figures~\ref{fig:etransform}, \ref{fig:braggedge} and \ref{fig:fastn}, there are several orders of magnitude between background and the peaks, and between peaks and non-elastic features. An elevated background could easily make the less intense features difficult or impossible to distinguish. Several effects may contribute to background. These include detector intrinsic background and $\gamma$-ray background. Sensitivity to fast neutrons described above also contributes to background. Due to the pulsed nature of a spallation source, fast neutrons due to the spallation process, arrive in a peak at a fixed time with respect to the elastic peak and other features. In practice, if the flux of fast neutrons is not reduced through shielding, it will be necessary to choose the incoming energy in such a way that the fast neutron pulse, or its tail, do not overlap with any features of interest that are of comparable or smaller amplitude. 

\subsubsection{Intrinsic Background}

Background due to the detector itself is independent of time and will contribute with a constant rate over the entire detector. The main contribution to this is due to $\alpha$ activity of detector material. In particular, aluminium, used as the substrate for the boron carbide and as the structural material of the grids can contain uranium, thorium and radium. This background causes energy deposition very similar to that of conversion products due to neutron interactions, and can therefore not be removed by a discriminator threshold. We have studied the effect of these $\alpha$-emitters in the earlier series of prototypes~\cite{cite:alpha}. In that work, two methods were successfully used in order to limit the background -- Ni-plating of the aluminium parts and use of ultra-pure aluminium. Background due to the detection of neutrons from cosmic rays and other non-facility sources cannot be distinguished from the $\alpha$ background in our measurements. Their contribution is, however, reduced by the shielded instrument cave that houses the detectors. 

\begin{figure}[tbp] 
\centering
\includegraphics[width=0.49\textwidth]{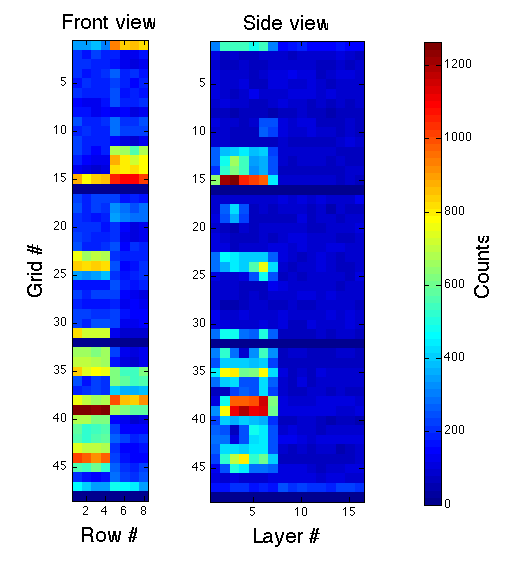}
\caption{An image of background counts measured during a 2-week beam outage. A front projection of the detector is shown on the left and a side projection on the right.}
\label{fig:intrinsic}
\end{figure}

The grids used in the MG.CNCS were made from Ni-plated aluminium, while the B$_4$C-coated blades were made of pure Al. In order to measure the intrinsic background, data was collected while the spallation source was off during a 2-week period of the target change. A background rate of 0.11~Hz in the entire Multi-Grid was measured on average during this time. In our previous prototype, tested at the IN6~\cite{cite:in6test}, where a background of 4.4~Hz was measured in similar conditions. Both prototypes had an equal number of grids. Considering the increased area and number of layers of the grids of MG.CNCS, the relative improvement is a factor 55. The rate in the $^3$He array in the same conditions was 0.4~Hz$/$m$^2$, or 0.08~Hz over the area equivalent to the MG.CNCS area. When examining the spatial distribution, it was found that some of the voxels show a much larger background than others, see figure~\ref{fig:intrinsic}. These are confined to the region where 0.5~$\mu$m coatings were used, suggesting that the affected blades came from the same batch. Approximately 23 grids showed one or more affected voxel. As the track of an $\alpha$ particle in Ar is several cm long, some of these 23 grids do not necessarily have any additional activity, but may be measuring $\alpha$s emitted from a neighbour grid. The number of affected voxels is consistent with a single sheet of regular aluminium used (44 blades were manufactured from every sheet). While it is not possible to verify this explanation, it is plausible, as other badges of blades were manufactured during the same period from regular aluminium, and a mistake may have been made while packing blades for different projects. Approximately 10\% of the voxel are affected to one degree or another. These contribute 55\% of the counts. We can therefore conclude that had issue not occurred, the background rate would be about a factor 2 lower, \emph{i.e.} 0.05~Hz.

\subsubsection{Gamma Background}

Gamma rays are common in the environment and are produced via most neutron absorption reactions on ambient nuclei. Gamma-rays from boron, cadmium and gadolinium -- materials used as neutron shielding -- are especially common at neutron facilities. Fortunately, $\gamma$-rays have a low interaction probability with the materials used in the Multi-Grid detector -- mainly Al and Ar. Furthermore, as described earlier~\cite{cite:khaplanov2}, $\gamma$-ray interactions result in energy deposits in the detector gas that are much smaller than the majority of the neutron conversions. 

\begin{figure}[tbp] 
\centering
\includegraphics[width=0.49\textwidth]{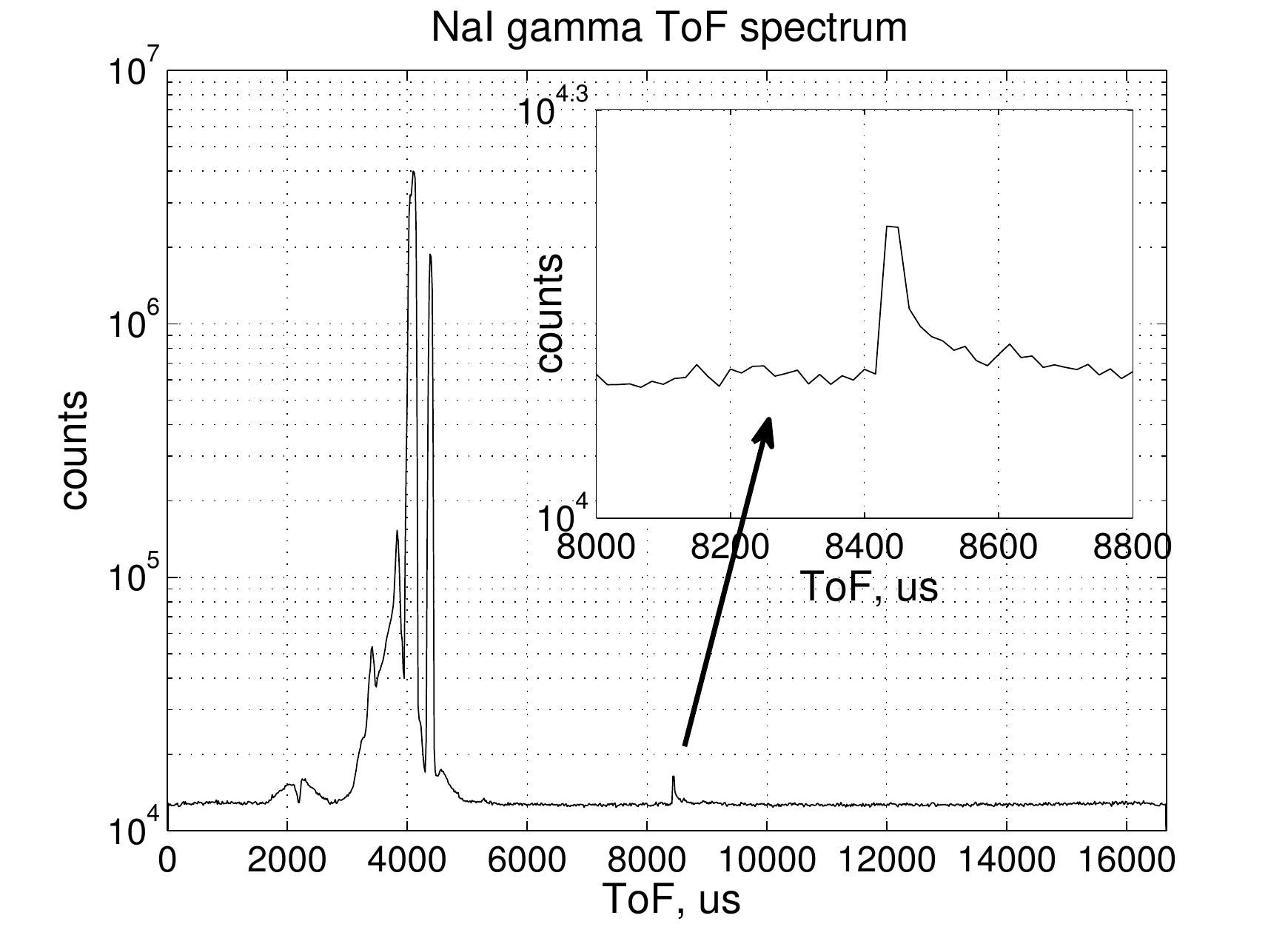}
\includegraphics[width=0.49\textwidth]{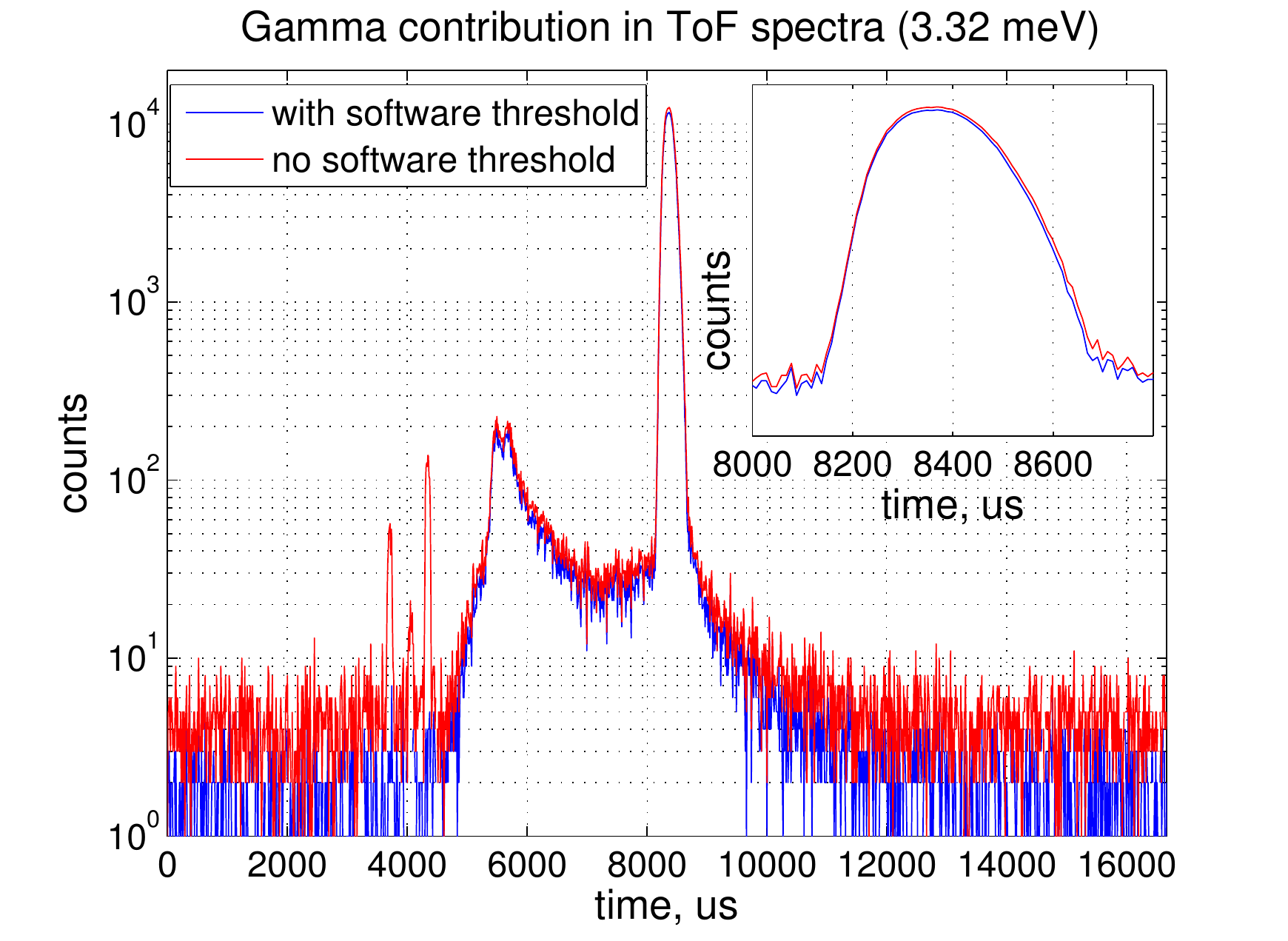}
\caption{\textbf{Left:} a $\gamma$-ray ToF spectrum measured by the NaI detector. \textbf{Right:} an Multi-Grid spectrum shown with and without a software threshold. The inserts magnify the region around the prompt and elastic peaks over the same time range, highlighting the difference in shape.}
\label{fig:gamma}
\end{figure}

A NaI scintillation detector was used to monitor the $\gamma$ flux. By connecting it to the MCA4, it was possible to obtain ToF spectra of $\gamma$-rays. One such spectrum is shown in figure~\ref{fig:gamma}. We see that there are several flashes of gamma rays just before the neutron pulse arrives at the sample (at the position where infinite energy gain neutrons would be found). This is expected, since a neutron pulse passing beam-forming elements will encounter absorbers, which emit $\gamma$-rays. The characteristic peak due to fast neutrons is also visible, but has a far lower amplitude. As the detector is shielded by boron, it is not expected to be affected by cold neutrons -- note that the elastic peak, which dominates neutron spectra, is absent here. Fast neutrons, however, will be able to penetrate the shielding and can be captured on both Na and I nuclei. These reactions emit $\gamma$-rays and conversion electrons and can thus be detected. The prompt fast neutron pulse is visible in figure~\ref{fig:gamma} at the middle of the ToF scale. It happens to overlap with the position of the elastic peak in this setting, however, it can be distinguished by its shape.

The low-energy threshold of the Multi-Grid detector has been set in the front-end electronics in such a way that the $\gamma$-ray signal is just barely visible. This allows to remove $\gamma$-rays using a software threshold while preserving the information on the level of $\gamma$ signals. The effect of the application of the threshold is shown in figure~\ref{fig:gamma} on the right. We see that the sharp gamma peaks are fully removed by the threshold, the time-independent background is slightly reduced, while other features are essentially unaffected. 

\section{Summary and Conclusion}

The testing performed so far with the Multi-Grid detector installed at CNCS has shown very encouraging results. We have tested several aspects of performance that can only be accessed in ToF measurements and confirmed the expectations from the initial characterisation using from test beams and source measurements. Furthermore, for each measurement, there is a corresponding $^3$He detector data set. This allows to compare the different properties of the two detector types as well as to isolate detector-related features from those related to the beam, sample environment and other parts of the instrument. 

\begin{description}
\item[Long-term operation] The detector has operated continuously for over 6 months with no intervention, apart from remote access to the data acquisition system and scheduled replacements of the gas supply. No deviations from steady state operations were detected; no adjustments to gains nor threshold settings have been necessary.

\item[Spectra] Comparisons of spectra in ToF and in the energy transfer scales show that the two detectors resolve all the same features. In the vanadium sample measurements in particular, elastic and inelastic signals, as well as the background level match closely. The effect of the prompt pulse is also visible in both. 

\item[Energy resolution] When converted into the energy transfer scale, the spectra from both detectors show equivalent energy resolution. This has been tested at several incoming energy resolution settings over the energy range from 0.76~meV to 80~meV and holds for all of the settings. 

\item[Efficiency] The counts in the Multi-Grid detector have been compared to those in $^3$He detectors, giving us a handle of relative efficiency. It varies from approximately 68\% at 80~meV to about 95\% at 1~meV. At 0.76~meV the counts in the Multi-Grid exceed those in $^3$He detector by 15\%. It is not clear whether this apparent increase in efficiency is real; it is not expected from analytical calculations of the efficiency for the Multi-Grid detector compared to tubes filled with $^3$He at 6~bar. Overall, we do not expect the measured efficiency difference to influence the quality of scientific data. 

\item[Bragg edge] High-resolution vanadium data has been collected just above and just below the Bragg edge energy in aluminium (3.742~meV). There are significant differences in the energy spectra. In particular, a number of side peaks appear on the energy-loss side of the elastic peak. These appear in both detectors, indicating that they are sample environment effects. Some increase in counts in the shoulder of the peak can be seen in the Multi-Grid detector. This will be further investigated using detailed simulations.

\item[Rate] The instantaneous rate capability was tested using a Bragg reflection from a single crystal. We have found that the Multi-Grid detector handles the high rate better. The position of neutron events and the rate variations as a function of the pulse time are reconstructed as expected. In case of the $^3$He detector, there is a loss of position resolution and a loss of count rate as the pulse reaches the peak. This can be understood due to the use of the charge division technique in the $^3$He tubes, whereas in the MG, charge division is not used, instead, each channel is read out individually and using a lower gas gain. The local count rate is also reduced due to a larger number of cells that contribute to each pixel. 

\item[Fast Neutrons] Neutrons that fail to thermalise in the moderator can reach the detector immediately after the proton pulse and constitute a background in measurements. They are seen in both detectors, although their counts in the Multi-Grid appear to be approximately half of $^3$He detectors. This may be due to the additional sensitivity to MeV neutrons via inelastic scattering on $^3$He, which has a greater cross section than the (n, p) reaction above 50~keV. 

\item[Background] Steps were taken to minimise the intrinsic background in the detector due to $\alpha$-emitting contaminants based on the previous experience with the Multi-Grid detectors. This was achieved using high-purity aluminium as well as Ni-plated parts. It is shown clearly that this approach works and this background is below the ambient background. 

\item[Gamma sensitivity] Background due to $\gamma$-rays is consistent with our earlier results. It is shown that with the correct choice of threshold, no $\gamma$ contribution can be detected over background.
\end{description}

The experience and measurements with the MG.CNCS demonstrator have shown that the detector is well suited for direct chopper spectrometers, such as CSPEC and T-REX under construction at the ESS. The current demonstrator has been optimised for a cold neutron spectrum, showing excellent performance at the neutron energies typically used at the CNCS. This matches closely the requirements of CSPEC. We are aiming to complete the validation of the Multi-Grid detector with a test of a demonstrator optimised for energies of around 100~meV on a thermal instrument. Additional work continues on simulations and the detailed understanding of neutron scattering effects within the detector and the instrument as a whole, as well as on data analysis procedures.

\acknowledgments

This work has been financed by the EU BrightnESS project, an H2020 grant agreement 676548. It used resources at the Spallation Neutron Source, a DOE Office of Science User Facility operated by the Oak Ridge National Laboratory. We would like to thank the personnel at the SNS for the support during the installation and measurements. We would like to thank M.~Janoschek (Los Alamos Nat. Lab.) and P.~Das (Ames Nat. Lab.) for the loan of the UGe$_2$ crystal.

\newpage
\section{Appendix}

The following figures~\ref{fig:de_spec1}, \ref{fig:de_spec2}, \ref{fig:de_spec3} show the energy transfer spectra in vanadium measurements performed over the course of this work. The FWHM of the peaks define the energy resolution data points shown in table~\ref{tab:eres} and figure~\ref{fig:eres}. The integral over the peak areas define the relative efficiency presented in table~\ref{tab:releff} and figure~\ref{fig:releff}.

\begin{figure}[h] 
\centering
\includegraphics[width=0.49\textwidth]{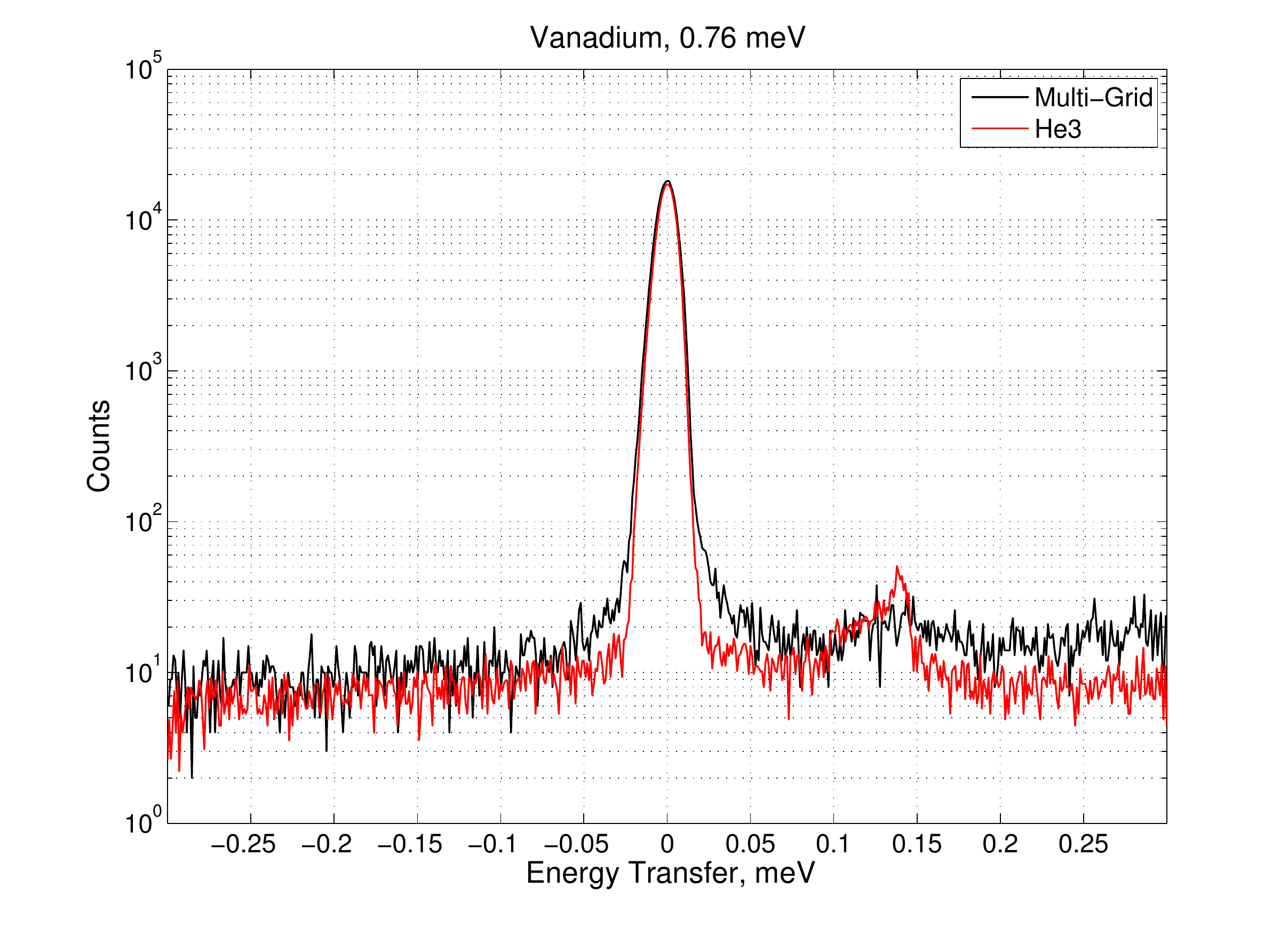}
\includegraphics[width=0.49\textwidth]{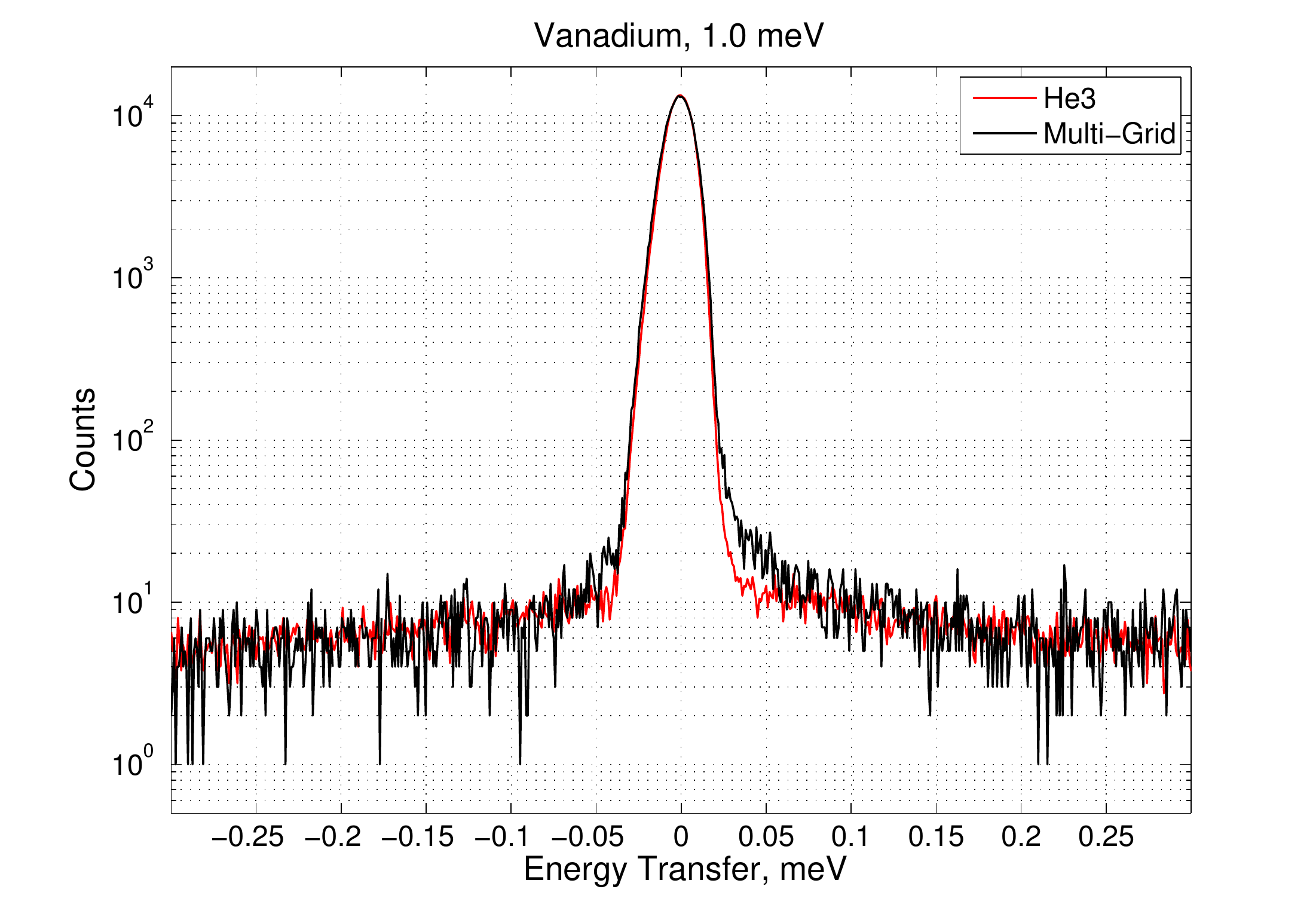}
\includegraphics[width=0.49\textwidth]{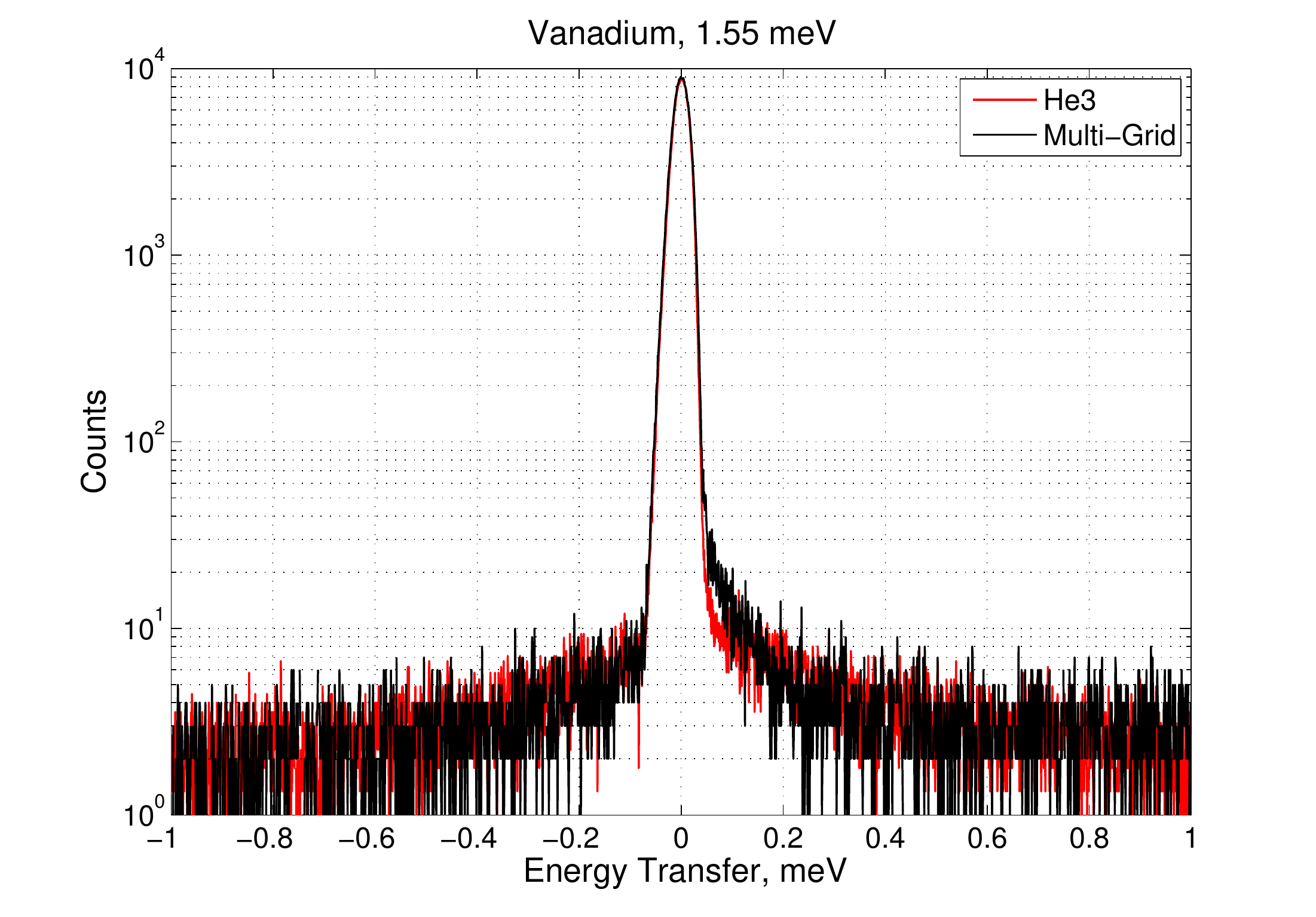}
\caption{The energy transfer spectra collected on July 12-13, in High Flux setting.}
\label{fig:de_spec1}
\end{figure}

\begin{figure}[tbp] 
\centering
\includegraphics[width=0.49\textwidth]{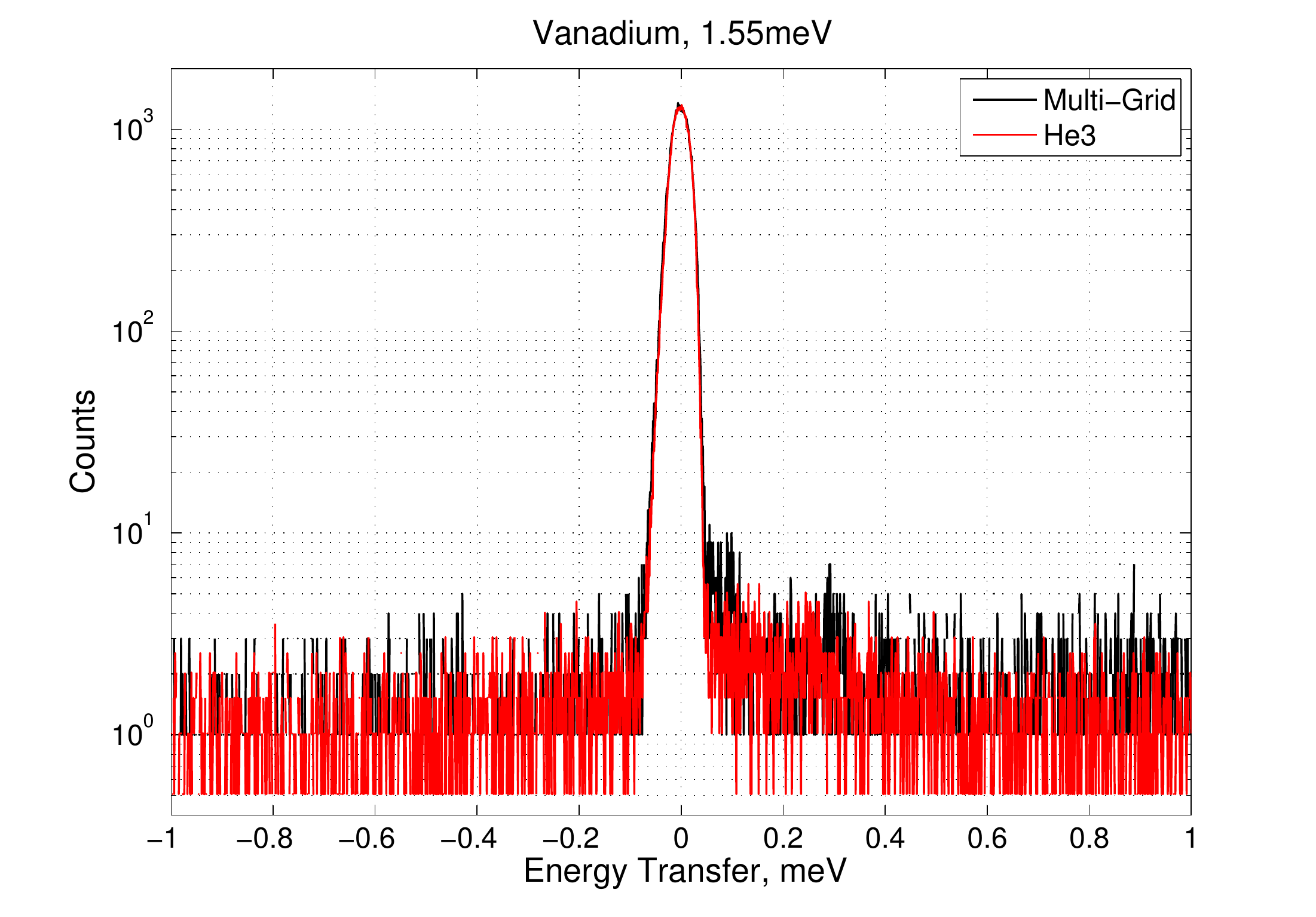}
\includegraphics[width=0.49\textwidth]{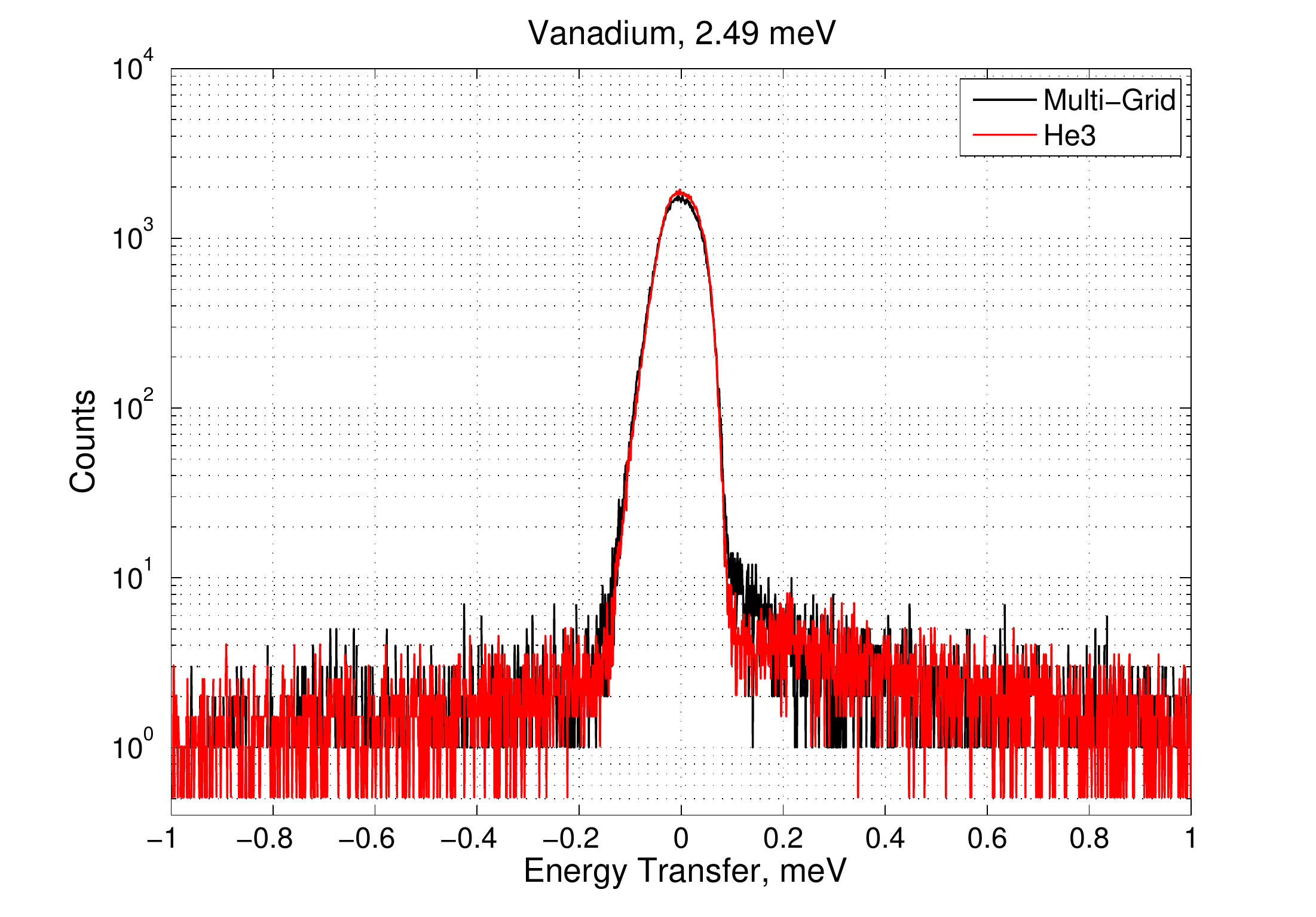}
\includegraphics[width=0.49\textwidth]{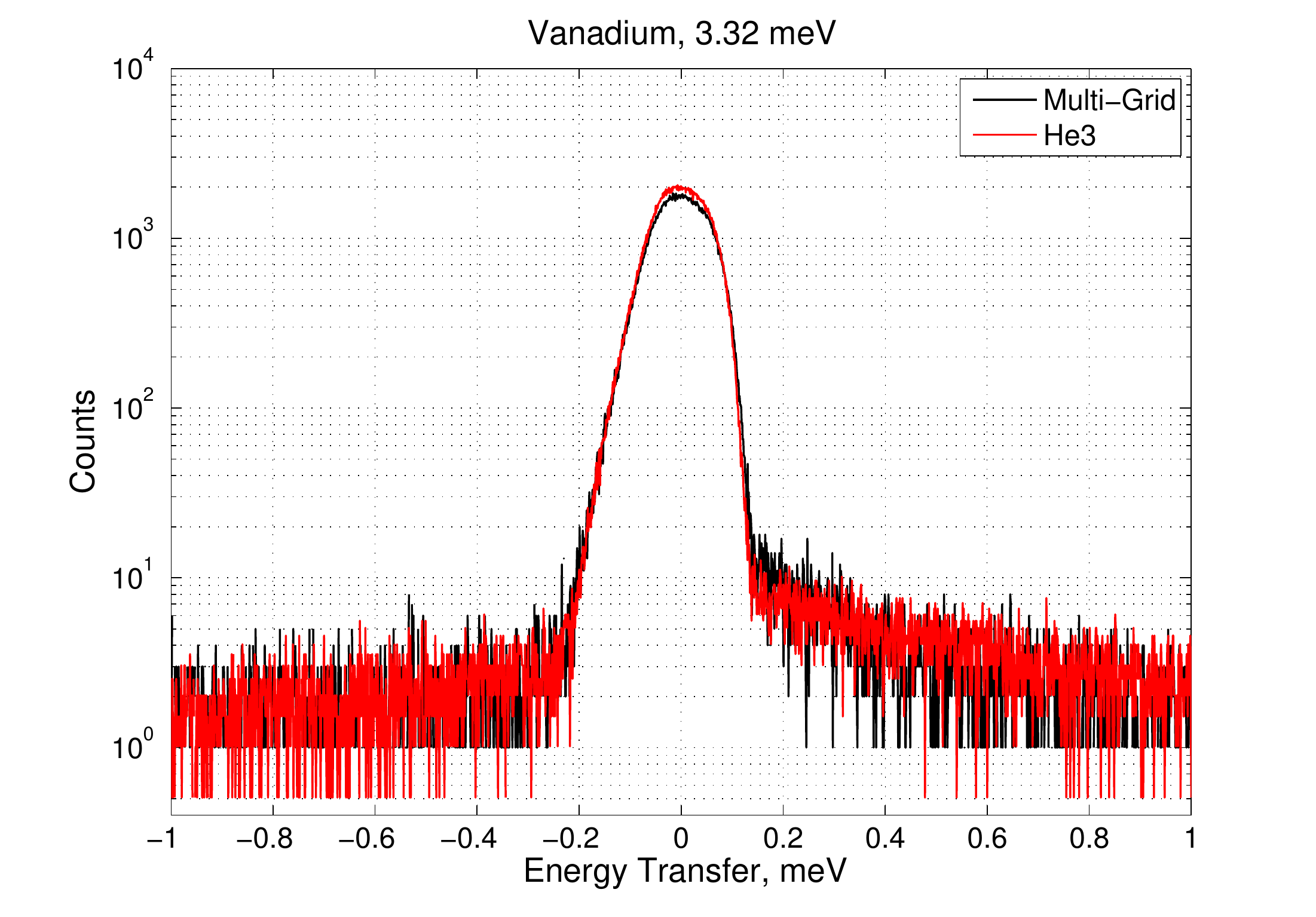}
\includegraphics[width=0.49\textwidth]{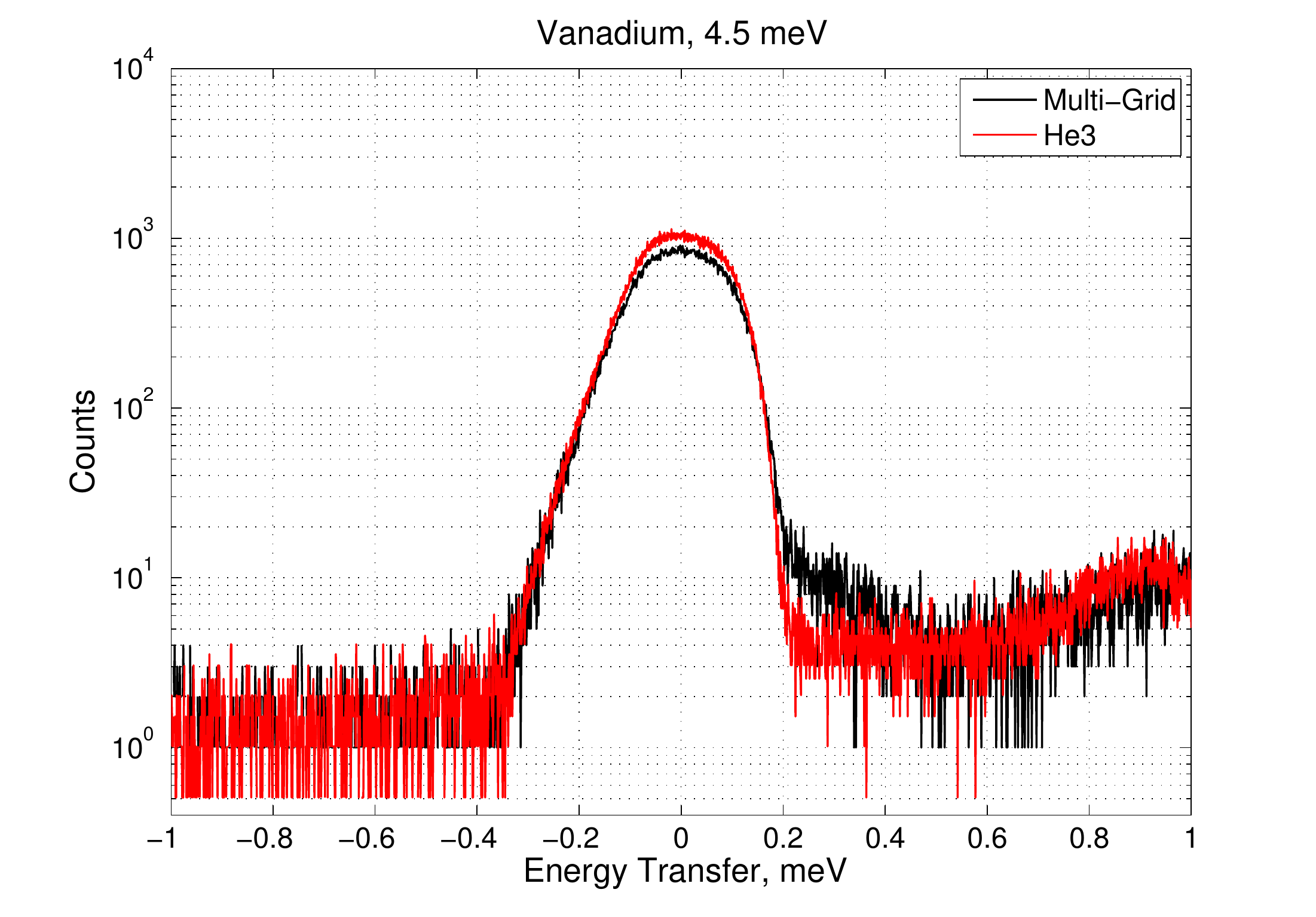}
\includegraphics[width=0.49\textwidth]{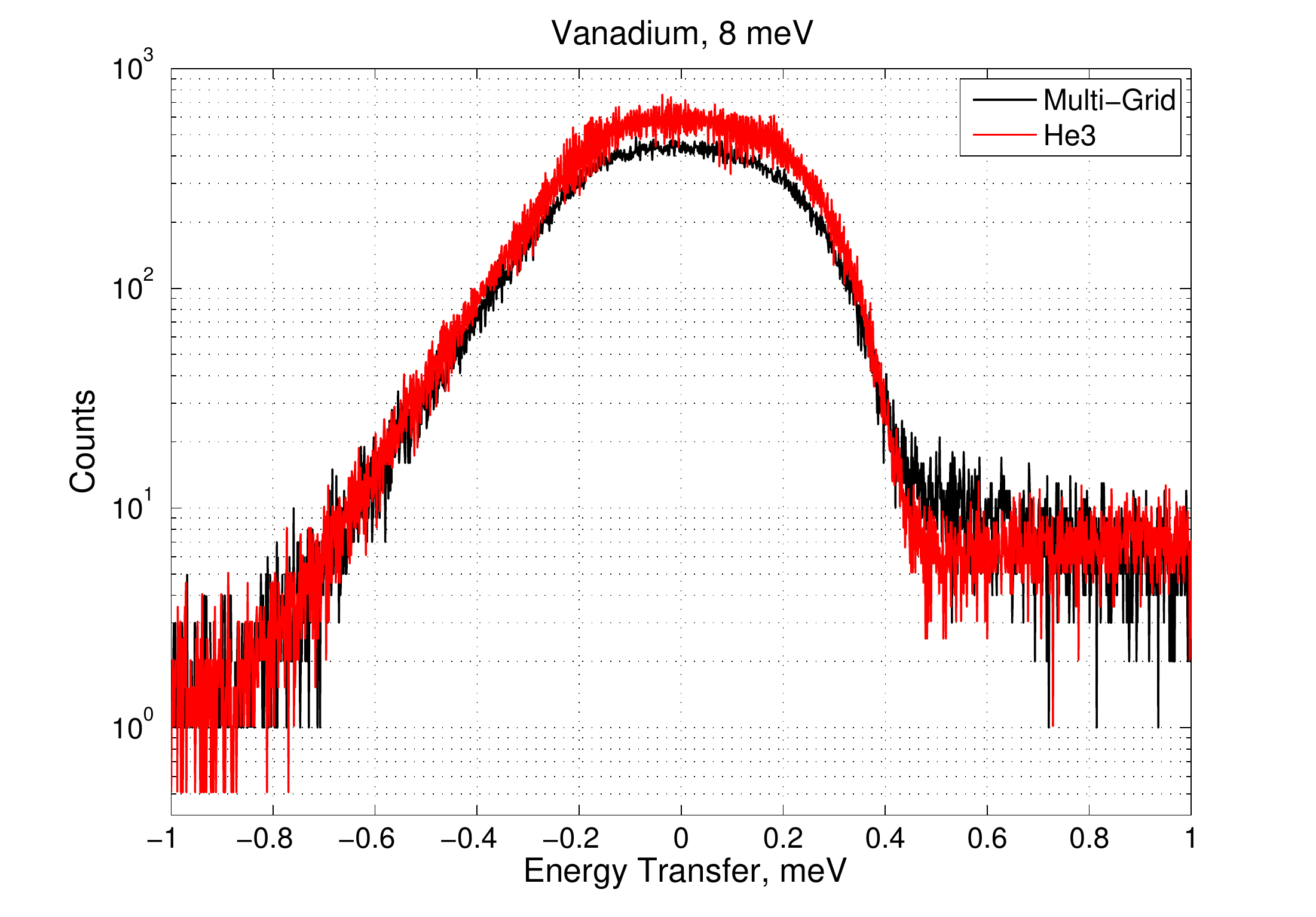}
\includegraphics[width=0.49\textwidth]{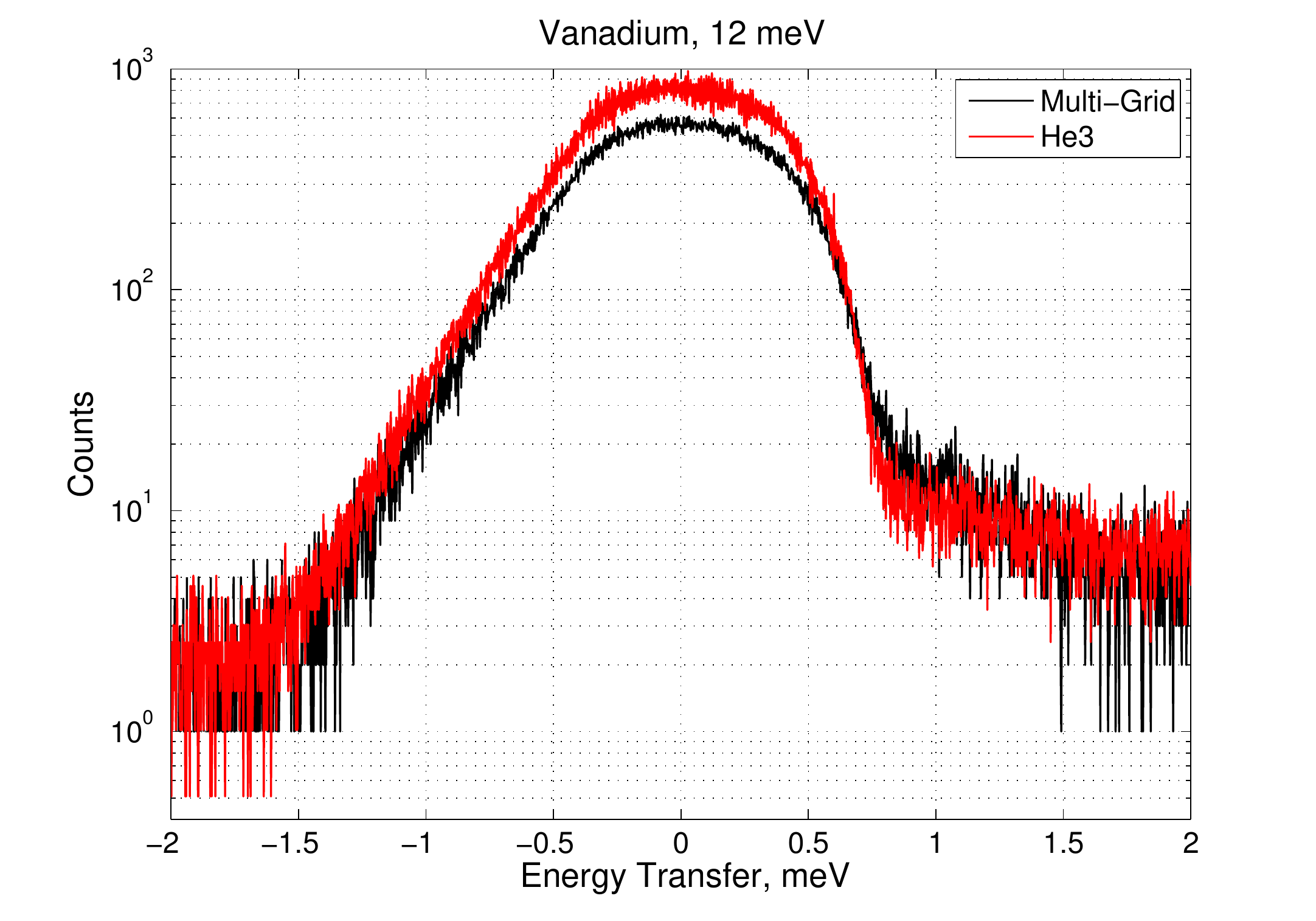}
\includegraphics[width=0.49\textwidth]{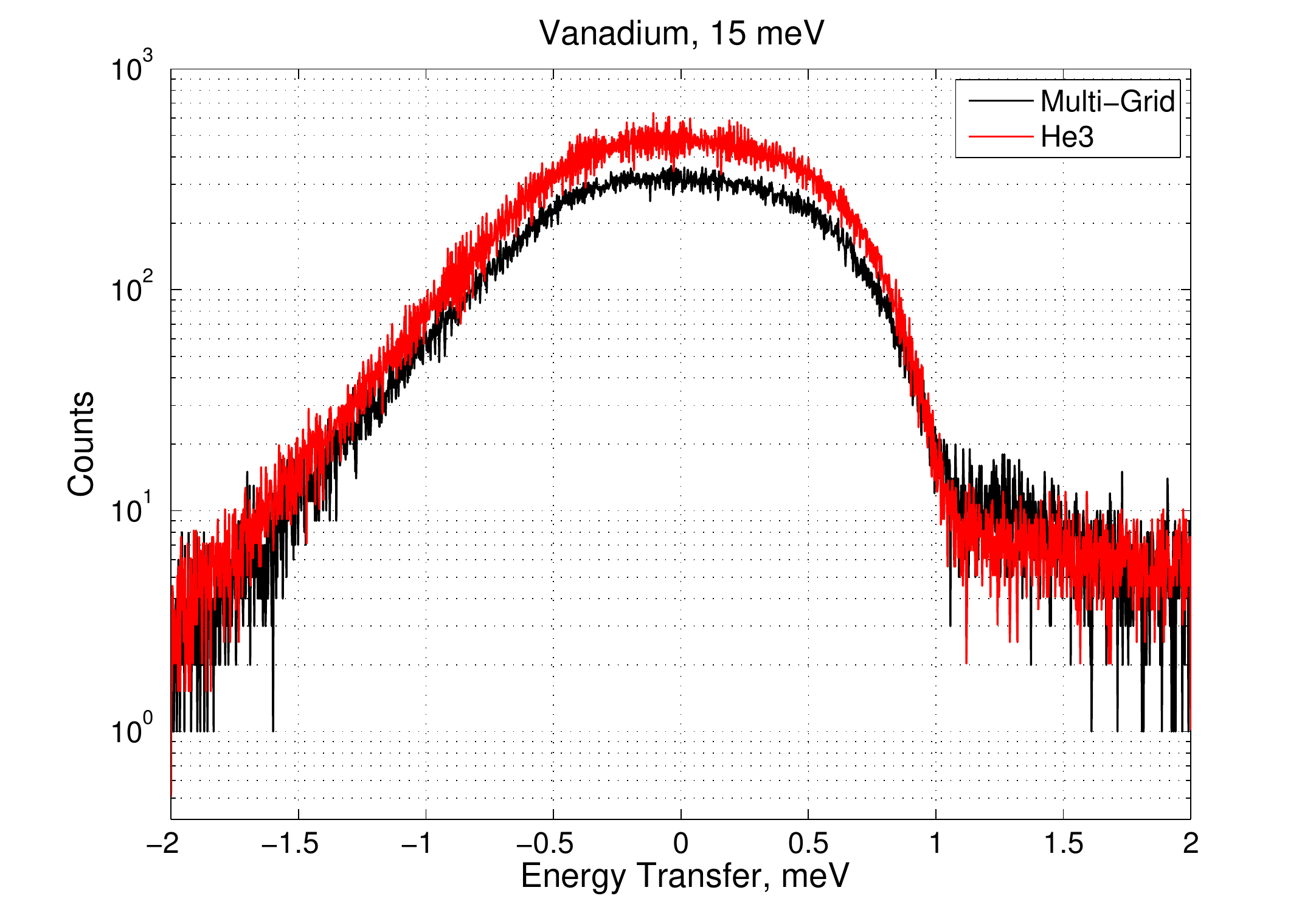}
\caption{The energy transfer spectra collected on July 27, in High Flux setting.}
\label{fig:de_spec2}
\end{figure}

\begin{figure}[tbp] 
\centering
\includegraphics[width=0.49\textwidth]{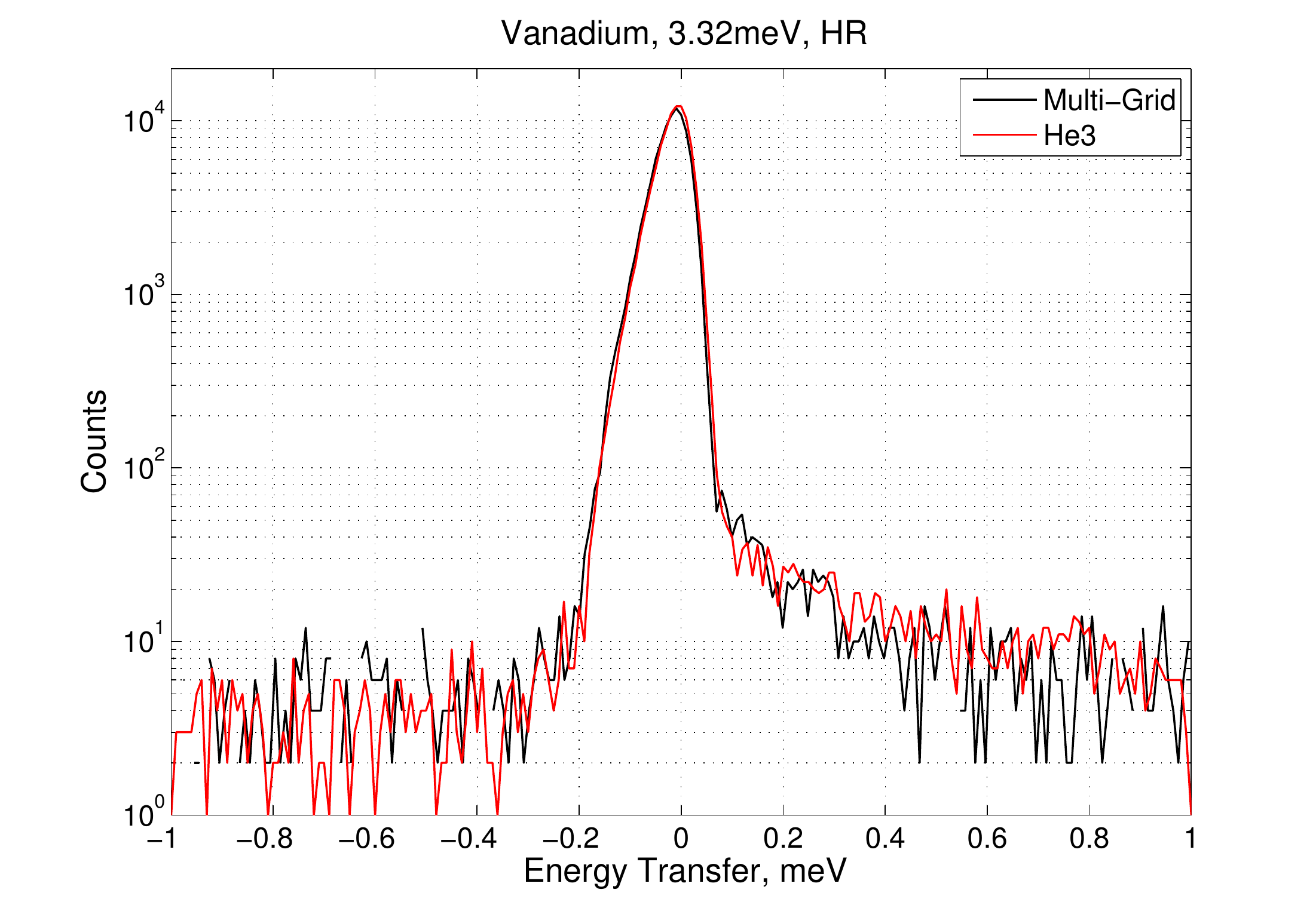}
\includegraphics[width=0.49\textwidth]{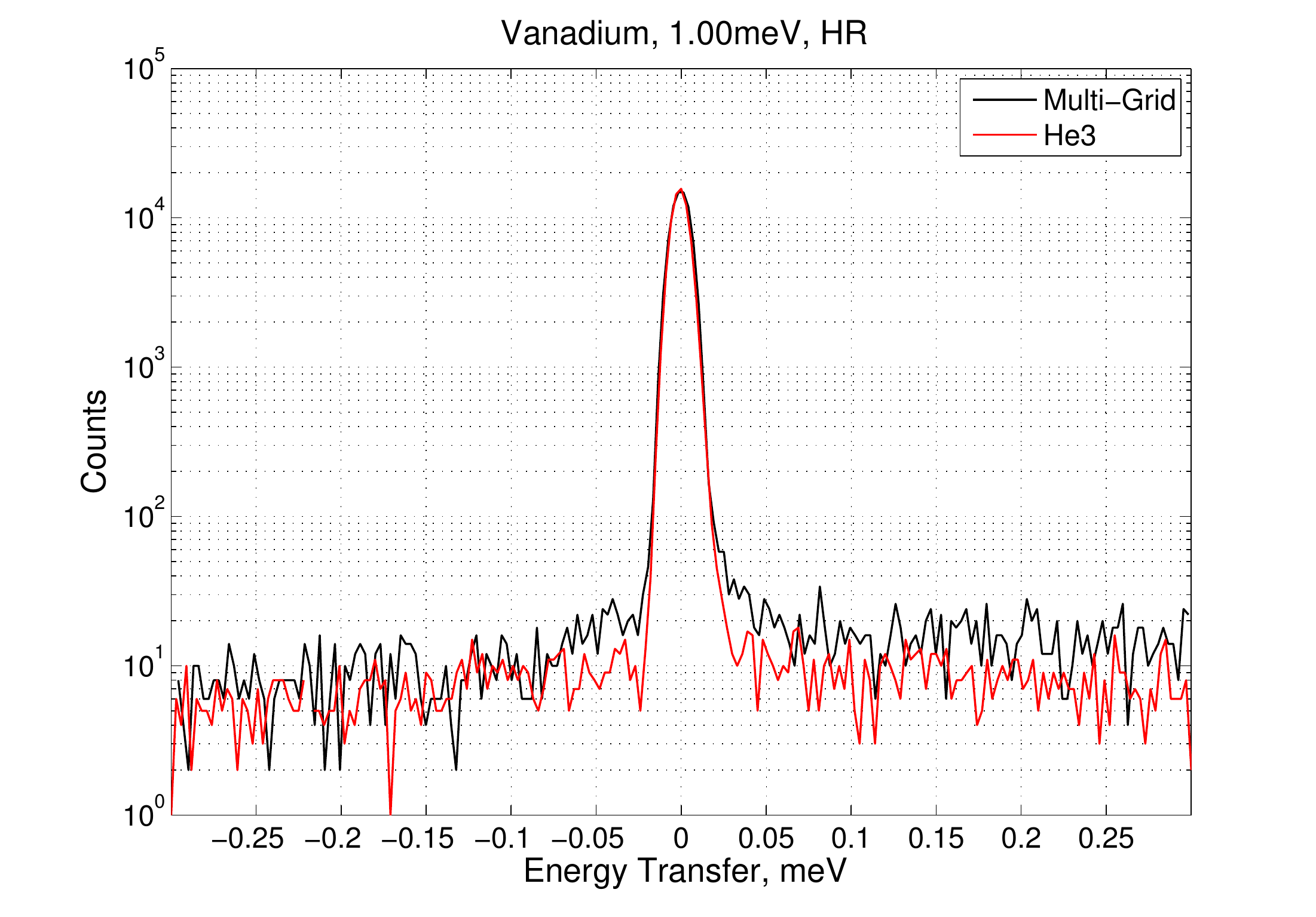}
\includegraphics[width=0.49\textwidth]{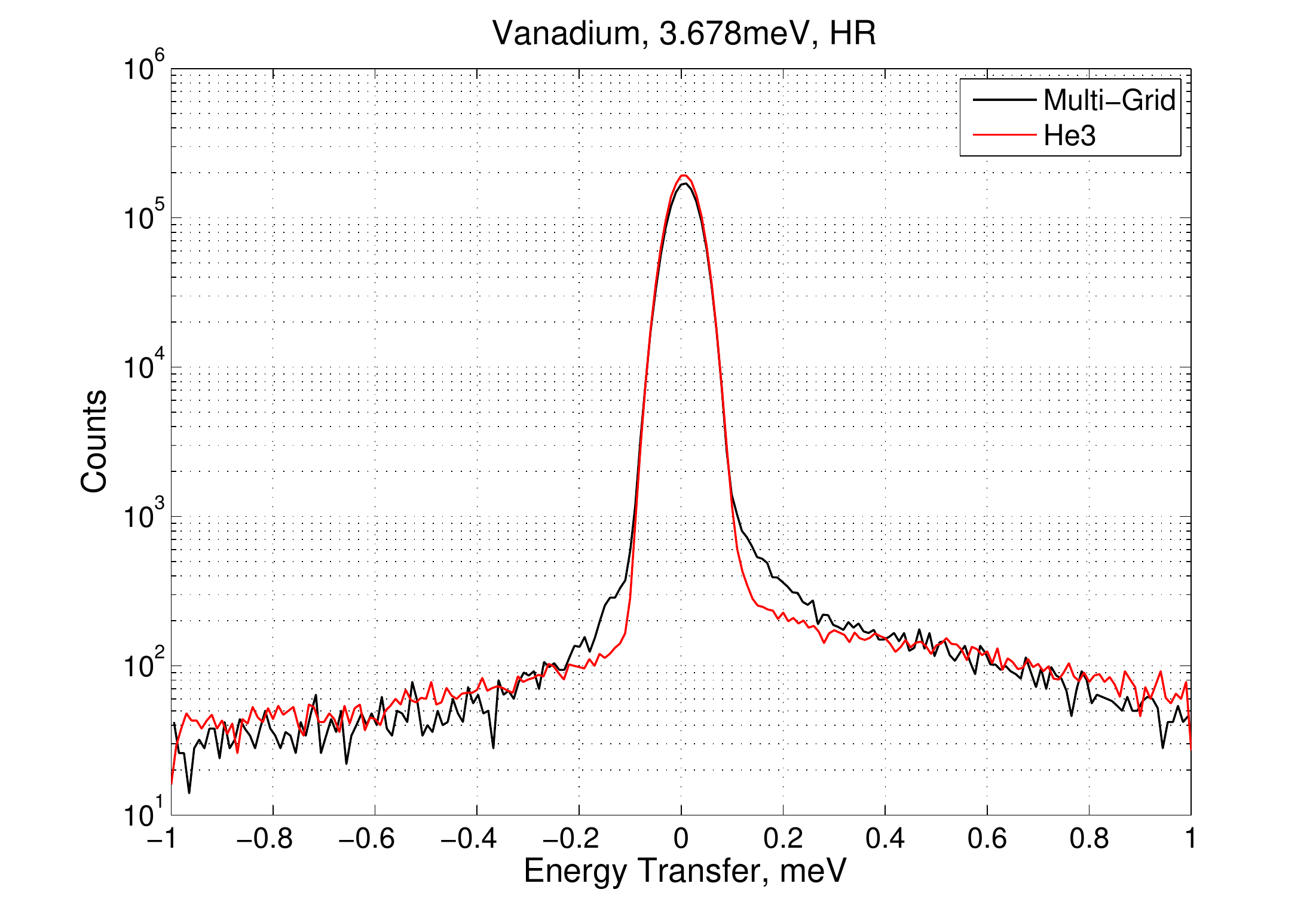}
\includegraphics[width=0.49\textwidth]{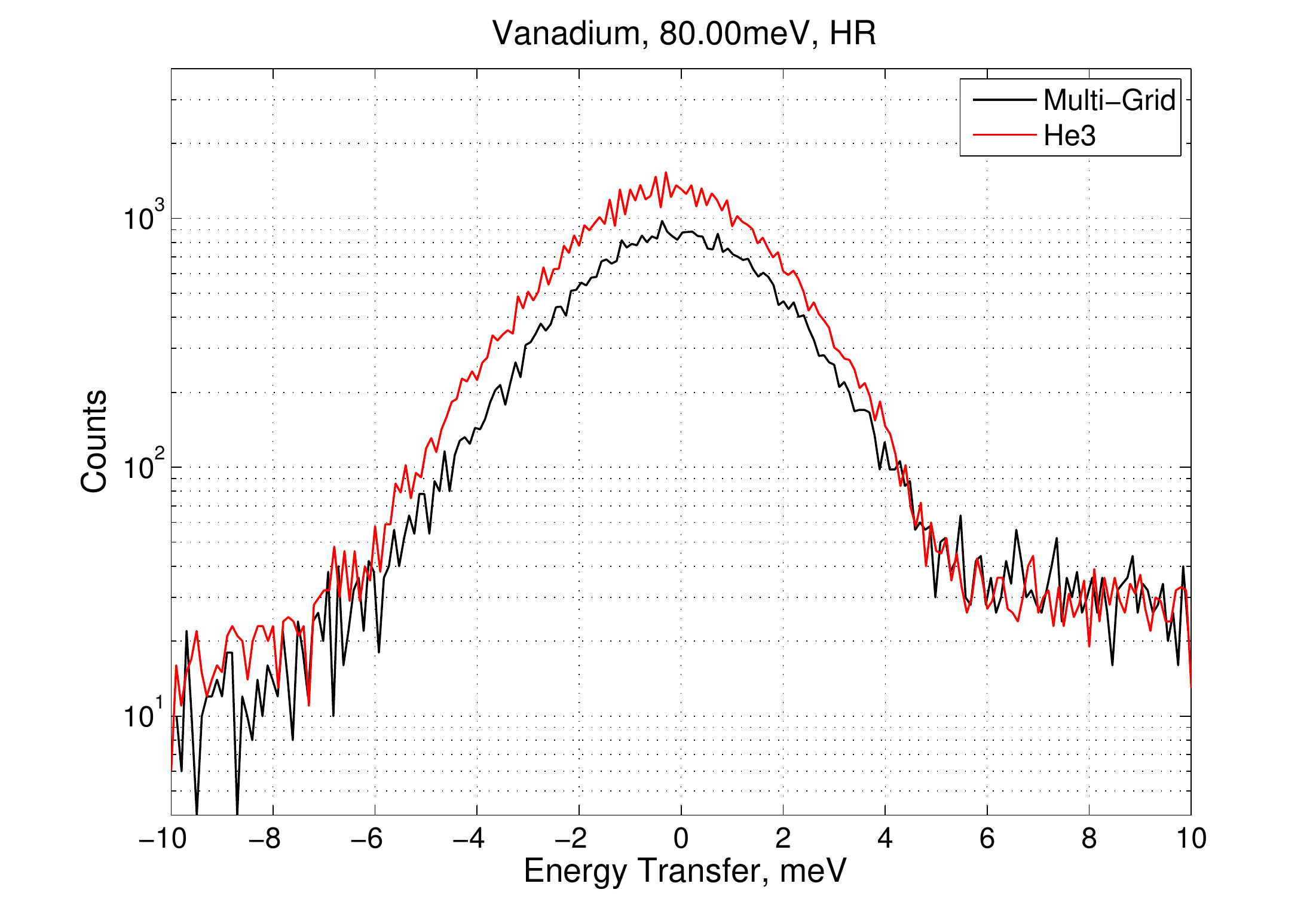}
\caption{The energy transfer spectra collected on September 07 (at 3.32~meV) and December 19 (at 1.00, 3.678 and 80.00~meV) in High Resolution settings.}
\label{fig:de_spec3}
\end{figure}

\end{document}